%% file: main.tex
\documentclass[twocolumn,twocolappendix,nofootinbib]{openjournal}

\usepackage[colorlinks,allcolors=blue]{hyperref}
\usepackage[utf8]{inputenc}
\usepackage[T1]{fontenc}

\usepackage{amsmath}
\usepackage{natbib}
\usepackage{newtxtext,newtxmath}
\usepackage{graphicx}
\usepackage{float}
\usepackage{svg}
\usepackage{booktabs}
\usepackage{listings}
\usepackage{tikz}
\usepackage{import}
\usepackage{fontawesome}

\makeatletter

\newcommand{\Rmnum}[1]{\expandafter\@slowromancap\romannumeral #1@}

\newcommand{\gsim}{\lower0.6ex\vbox{\hbox{$\buildrel{\textstyle >}\over{\sim}\ $}}}

\usepackage{color}

\newcommand{\xw}[1]{\textcolor{black}{#1}}

\def\mpch{h^{-1}{\rm Mpc}}
\def\hmpc{h{\rm Mpc}^{-1}}

\def\invhmpc{\;h\;{\rm Mpc}^{-1}}

\def\kms{\, {\rm km}\, {\rm s}^{-1}}

\def\hmsun{{h^{-1} M_{\odot}}}
\def\msun{\, M_{\odot}}

\begin{document}
\journalinfo{The Open Journal of Astrophysics}

\title[Emulator for Super Resolution Model]
{AI-assisted super-resolution cosmological simulations IV: An emulator for deterministic realizations}

\author{
Xiaowen Zhang$^{1}$,
Patrick Lachance$^{1}$,
Ankita Dasgupta$^{2}$,
Rupert A.~C. Croft$^{1}$,
Tiziana Di Matteo$^{1}$,
Yueying Ni$^{3}$,\\
Simeon Bird$^{4}$ and
Yin Li$^{5}$ 
}
\thanks{Email:xiaowen4@andrew.cmu.edu}
\affiliation{
$^1$ McWilliams Center for Cosmology, Department of Physics, Carnegie Mellon University, Pittsburgh, PA 15213 \\
$^2$ Department of Astronomy and Astrophysics, Pennsylvania State University, University Park, PA 16802, USA \\
$^3$ Harvard-Smithsonian Center for Astrophysics, 60 Garden Street, Cambridge, MA 02138, USA \\
$^4$ Department of Physics and Astronomy, University of California Riverside \\
$^5$ Department of Mathematics and Theory, Peng Cheng Laboratory, Shenzhen, Guangdong 518066, China
}

\begin{abstract}
Super-resolution (SR) models in cosmological simulations use deep learning (DL) to rapidly enhance low-resolution (LR) runs with statistically correct fine details. These models preserves large-scale structures by conditioning on an LR version of the simulation. On smaller scales, the generative process is inherently stochastic, producing multiple possible SR realizations with distinct small-scale structures. Validation of \xw{reconstructed SR runs from LR simulations requires ensuring that specific statistics of interest are accurately reproduced by comparing SR outputs with target high resolution (HR) runs}. In this study, we develop an emulator designed to reproduce the small-scale structures of \xw{target HR} simulation with high fidelity. By processing an SR realization alongside the high-resolution initial condition (HRIC), we transform the SR output to emulate the result of a full simulation with that HRIC. \xw{By comparing various metrics, from visualization to individual halo measurements, we demonstrate that the emulated SR runs closely align with the target HR simulation, even at length scales an order of magnitude smaller than the corresponding LR run}. These results show the potential of this method for efficiently generating accurate simulations and mock observations for large galaxy surveys.

\end{abstract}

\maketitle

\input{Sec1_Introduction.tex}
\input{Sec2_Method.tex}
\input{Sec3_Result1_visual}

\input{Sec3_Result2_spectrum}
\input{Sec3_Result3_halo}

\input{Sec4_Discussion.tex}

\input{Sec5_Conclusion.tex}

\section*{Data Availability}
The model framework is based on \texttt{map2map} \url{https://github.com/eelregit/map2map}. This PyTorch-based framework is a general-purpose tool for transforming field data. The trained model weights and the pipeline to generate the fields will be available at (\url{https://github.com/xwzhang98/SREmulator}). Training and test sets data generated in this work will be shared on reasonable request to the corresponding author.

\section*{Acknowledgements}
This research is part of the Frontera computing project at the Texas Advanced Computing Center. Frontera is made possible by NSF award OAC-1818253.
TDM acknowledges funding from NASA ATP 19-ATP19-0084 and 80NSSC20K0519.
TDM and RACC also acknowledge funding from NASA ATP 80NSSC18K101, and NASA ATP NNX17AK56G.
SB was supported by NASA ATP 80NSSC22K1897. 
This work and AD's participation was also supported by the NSF AI Institute: Physics of the Future, NSF PHY-2020295.
The computational results presented have been achieved using Vera at Pittsburgh Supercomputing Center (PSC).
This work also used Anvil at Purdue University through allocation PHY240032 from the Advanced Cyberinfrastructure Coordination Ecosystem: Services \& Support (ACCESS) program, which is supported by National Science Foundation grants 2138259, 2138286, 2138307, 2137603, and 2138296.
We also acknowledge the code packages used in this work:
The model framework is based on \texttt{map2map} \url{https://github.com/eelregit/map2map}. The simulations for training and testing is run with \texttt{MP-Gadget} (\url{https://github.com/MP-Gadget/MP-Gadget}). Visualization in this work is performed with open source code \texttt{gaepsi2} (\url{https://github.com/rainwoodman/gaepsi2}) and \texttt{plotly} \citep{plotly}. Data and catalog analysis in this work is performed with open-source software \texttt{PyTorch}\citep{pytorch}, \texttt{nbodykit}\citep{Hand2018} and \texttt{ytree}\citep{ytree}.

\appendix
\input{Sec6_Appendix}

\bibliographystyle{mnras}
\bibliography{bib_files/physics, bib_files/ml, bib_files/software_package}
\end{document}

%% file: Sec1_Introduction.tex
\section{Introduction}
\label{section1:introduction}

Cosmological N-body simulations have served as powerful computational tools since the 1960s \citep{nbodyfirst, nbodysecond} for solving the nonlinear dynamics intrinsic to cosmic structure formation. High-resolution (HR) simulations employing state-of-the-art, massively parallel codes, such as \texttt{MP-Gadget}, evolve a large number of particles under gravity, tracking the formation of structures on the order of galaxy size. Hydrodynamic simulations, such as Illustris-TNG \citep{IllustrisTNG}, SIMBA \citep{Simba} and ASTRID \citep{astrid, astrid2}, which model the baryonic matter as a separate component, require even more computational investment compared to gravity-only N-body simulations. These simulations require substantial time and computational resources, forcing researchers to balance resolution and simulation volume. As a result, they must often choose between reducing resolution to meet computational constraints or limiting the simulation volume to achieve finer scales. For example, the small volume, high-resolution FIRE suite \citep{fire} while ABACUS summit \citep{abacus} runs are optimized for large-scale structure.

In recent years, Machine Learning (ML), particularly Deep Learning (DL), has emerged as a powerful tool for physicists \citep{MLincosmo}. ML models have been employed to predict various baryonic properties from dark matter-only simulations, including the galaxy distribution \citep{Modi2018, zhang2019}, the tSZ effect \citep{Troster2019}, the 21 cm emission distribution \citep{wadekar2020hinet}, and stellar maps \citep{Dai2021}. Predictions of nonlinear structure from linear cosmological initial conditions ~\citep{he2019learning, berger2019, Bernardini2020, Renan20} and from the dark matter density field \citep{KodiRamanah2020} have proved that DL based Neural Networks (NNs) can capture highly nonlinear structures.

Training large NNs requires extensive, high-quality datasets. An example of such a simulated dataset is CAMELS~\citep[Cosmology and Astrophysics with MachinE Learning Simulations,][]{Villanavarro2020b}, a
large-scale simulation project with a training set of over 4000 hydrodynamical simulations run with different hydrodynamic solvers and subgrid models for galaxy formation. The large size and comprehensive range of parameter variations in this dataset enables investigations using machine learning techniques into the interplay of baryonic effects and cosmology.

\xw{To reduce computational demand and also achieve result comparable in accuracy to full HR N-body simulations}, researchers have developed Super-Resolution (SR) techniques to lower resolution cosmological simulations (see e.g., \citealt{AI1} and other papers in this series). SR in general describes the situation where information is added below the resolution scale of initially lower resolution data such as a picture (e.g., in \citealt{unetpercep}) to create an HR version. However, SR remains a challenging and ill-posed problem, because an infinite number of HR results can be downsampled to the same LR input, meaning the SR model learns a conditional probabilistic distribution of HR simulations based on a single LR simulation. This `one-to-many' regime can be useful when many simulations or mock observations are needed for accurate and precise cosmological parameter inference.

Generative Adversarial Network (GAN) \citep{goodfellow2014GAN} architectures have already proven to be effective in SR tasks \citep{wang2020} and have been applied to the generation of 2D images of cosmic webs \citep{Rodrguez2018}, 3D matter density fields \citep{Perraudin2019} and cosmological mass maps \citep{Perraudin2020}. GANs consist of two separate NNs: a generator, which generates candidates, and a discriminator, which provides feedback by comparing real data with generated candidates. In previous work, an SR framework which applies a GAN approach directly to $N$-body simulation particles has been explored, covering from displacement only \citep{AI1} to full 6-D phase space \citep{AI2}, and which can also be conditioned on redshift \citep{Ai3} (hereafter Paper I, Paper II and Paper III). The output of this framework can be treated and analyzed in the same manner as full $N$-body simulation data. The same framework has also been used for fuzzy dark matter cosmology \citep{fuzzy-sr}. In Paper I, II and III, we trained our GAN based framework to perform a stochastic mapping from LR input to SR output, where stochasticity is injected during the generation process. These models aim to learn a probability distribution of HR outputs conditioned on LR inputs. By sampling this distribution using a special technique, as described in \cite{Ai3}, we can carefully control the randomness in the SR field, allowing us to generate simulations with outputs at any desired redshift. Instead of checking that summary statistics of the SR model are within expected bounds, an SR run that reproduces the individual structures of the HR run would (at least down to some length or mass scale) be a valid cosmological simulation, one that could be used to compute any summary statistic. This is one of the goals of this paper. Recently, denoising diffusion models \citep{diffusion2015, diffusion_ho} have become state-of-the-art, outperforming GANs for some image generation tasks and also being used in SR. It has been shown that super-resolving a 3D density field conditioned on an LR density field \citep{Rouhiainen2023} using a diffusion model has comparable performance to GAN results and does not suffer from unstable training and mode collapse. Several approaches employing diffusion models demonstrate the potential of this method, such as in 2D galaxy images \citep{Smith2022} and in 2D simulations \citep{diffusion-sr} using filter-boosted tricks. However, their work is limited to 2D cases only.

In this work, we introduce an emulator framework which can adjust the SR outputs to their HR counterparts by conditioning on specific high-resolution initial conditions (HRICs). These specific HRICs have the same large-scale modes as the LR $N$-body simulations that are used to generate SR outputs. Previous studies have also utilized emulators, such as \cite{SR_emulator}, which combines LR and HRICs for super-resolution tasks on 3D density fields. Similar to our approach, their model does not incorporate stochasticity. Furthermore, \cite{field_level_emulator} uses a neural network to map the linear approximation at $z = 0$ to full $N$-body simulations under various cosmological parameters, using a Lagrangian description. To achieve accurate SR emulation of the HR simulations at small scales, we find that using adversarial training techniques is crucial for correcting small-scale discrepancies. In this study, we explore emulators trained with this technique (hereafter referred to as "Emulator"), and compare them to our previous work in \cite{Ai3} (hereafter referred to as "SR").

The paper is structured as follows. In Section \ref{section2:Method}, we discuss the generation of data sets and summarize the details of our neural networks and the training process. In Section \ref{section3:Results} we show the performance of our emulator, specifically visual validation in \ref{subsection3.1: Visual}, Fourier mode cross-correlations in \ref{subsection3.2: Power Spectra} and halo catalog in \ref{subsection3.3: halo-catalog}. In Section \ref{section4:Discussion}, we discuss our findings, and we summarize our results and conclude in Section \ref{section5:Conclusion}. We have also included the results of the training without discriminator guidance (hereafter referred to as Emulator w/o DG) in the appendix \ref{appendix}.

%% file: Sec2_Method.tex
\section{Method}
\label{section2:Method}

\subsection{Dataset}
\label{subsection2.1:dataset}

The $N$-body simulation is a widely used tool for understanding the nonlinear evolution of cosmic structures. Given a fixed set of cosmological parameters \citep[in this work, we use the WMAP9 cosmology, as prescribed by][]{hinshaw13}, this method involves dividing the mass distribution into $N$ equal-mass particles. We run both LR and HR simulations using \texttt{MP-Gadget}\footnote{\url{https://github.com/MP-Gadget/MP-Gadget}}, a highly parallel code that solves the gravitational force using the TreePM method \citep{bagla02}. Long-range forces are calculated on a particle mesh in Fourier space, and short-range forces are computed through a hierarchical tree structure. Our LR and HR simulation tasks follow the Lagrangian description, where particles are tracers of both displacement and velocity fields. The displacement of an individual particle is:
\begin{align}
    \mathbf{d}_i = \mathbf{x}_i - \mathbf{q}_i \label{eq:dis}
\end{align}
where $\mathbf{d}_i$ is the displacement of $i$ th particle, $\mathbf{x}_i$ is the current position and $\mathbf{q}_i$ is its initial position on the uniform grid. Similarly, the velocity of individual particle is :
\begin{align}
    \mathbf{v}_i = \dot{\mathbf{x}_i} = \dot{\mathbf{d}_i}, \label{eq:vel}
\end{align}
where the dots denote time derivatives.
The initial conditions for our simulations are generated at $z = 99$ using a power spectrum calculated by CLASS \citep{Lesgourgues2011}. First-order Lagrangian perturbation theory \citep{Zel'dovich1970, Crocce2006} is used to initialize particle positions and velocities. The definition of initial displacement $\mathbf{d}_{init, i}$ and velocity $\mathbf{v}_{init, i}$ for particles is similar to Eq. \ref{eq:dis} and \ref{eq:vel}. Cosmological parameters follow the WMAP9 cosmology \citep{hinshaw13}, with matter density $\Omega _{\rm m} = 0.2814$, dark energy density $\Omega_{\Lambda} = 0.7186$, baryon density $\Omega {\rm b} = 0.0464$, power spectrum normalization $\sigma_{8} = 0.82$, spectral index $n_{s} = 0.971$, and Hubble parameter $h = 0.697$. The mean spatial separation of the dark matter particles is used to determine the gravitational softening length, which is set to $1/30$ of this value. Our training dataset is comprised of 8 dark matter only simulations, each with LR and HR simulation pairs. Each LR-HR simulation pair has $64^3$ and $512^3$ particles, respectively, with box size $100 \mpch$. The mass resolution is $m_{\mathrm{DM}} = 2.98 \times 10^{11} \hmsun$ for LR, and $m_{\mathrm{DM}} = 5.8 \times 10^{8} \hmsun$ for HR. Each pair has a unique random seed for its initial condition, so the initial conditions differ for eight simulations. In addition, we also create a test dataset with 3 LR and HR simulation pairs. Each pair consists of 60 snapshots from $z = 1$ to $z = 0$, with the same cosmological parameters as the training set but different initial condition random seeds. \xw{During data pre-processing, the particle displacement and velocity fields are divided into cubical chunks. These fields are then concatenated channel-wise, as they share the same spatial resolution. The same procedure is applied to the HRICs.}

\subsection{Model Architecture and General Training Process}
\label{subsection2.2:unet}

\begin{figure*}
\centering
\includegraphics[width=\textwidth]{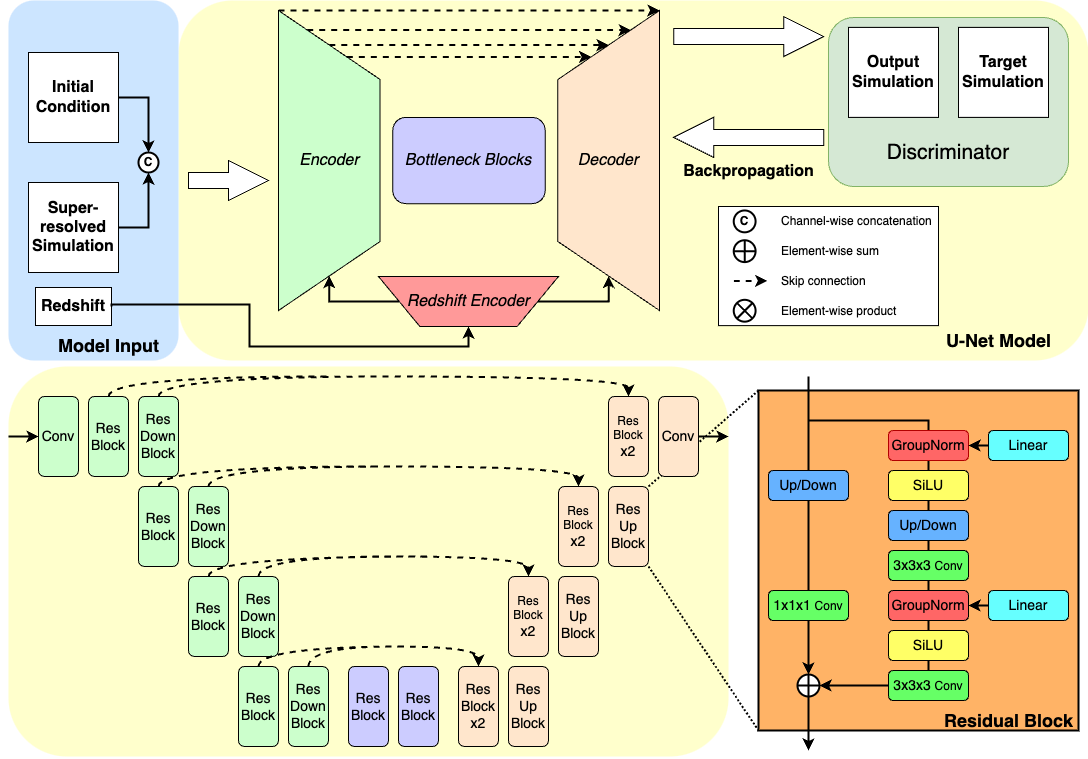}
\caption{
\textit{\textbf{Upper Panel :}}
A schematic representation of the main components of the model and the overall training process.
\textit{\textbf{Lower Panel :}}
The architecture of our U-Net model and the detailed structure of the Residual Blocks. The \textit{Res up Block} and \textit{Res Down Block} represent the up-sampling and down-sampling components, respectively, while the \textit{Res Blocks} do not include up-sampling or down-sampling operations. For a more detailed description, refer to Section \ref{subsection2.2:unet}.
}
\label{fig:MLStructure}
\end{figure*}

\xw{Our model employs a U-Net architecture, named for its U-shaped design, originally developed for image segmentation \citep{Unet} and adapted here for 3D tasks. U-Nets have also been applied to super-resolution and style transfer tasks \citep{unetpercep}, as well as to image to image translation, image uncropping, and more \cite[see discussion in][for a review]{palette}. Recent diffusion models \citep{diffusion2015, diffusion_ho} also use U-Nets for denoising. The U-Net architecture consists of an encoder and a decoder connected by a bottleneck part. The U-Net architecture consists of an encoder, a bottleneck, and a decoder. The encoder progressively downsamples the input towards the bottleneck, while the decoder upsamples the data back to its original resolution. "Skip-connections" link the encoder and decoder at corresponding resolutions, preserving spatial information throughout the network.}

\xw{The upper panel of Figure \ref{fig:MLStructure} illustrates the overall layout of our architecture and training process. The model's redshift encoder consists of a multi-layer perceptron designed to extract redshift information, enabling the generation of outputs at multiple epochs. The initial condition and SR realization are concatenated channel-wise and then fed into the U-Net model. The output simulation is then compared to the target HR simulation using trainable discriminator. During training, backpropagation is performed using Pytorch, with a combined loss function employed to optimize particle-level details.}
\xw{The lower panel of Figure \ref{fig:MLStructure} details each component's role}. The encoder part extracts features from the input and downsamples them towards the bottleneck. This forces the model to learn the key information from the input and embed it into a low-dimensional space. Starting from the input resolution, the decoder downsamples the learned feature until it reaches the bottleneck blocks. Each decoder block at the same resolution consists of two residual blocks. We follow BigGAN's \citep{biggan} residual block structure since it has been proven to have good performance and can be scaled to different depths. All residual blocks contain two branches, one ``skip'' branch with a 3D size 1 convolutional block, the other ``main'' branch, which first normalizes the input with Group Normalization \citep{groupnorm}, then passes through a Sigmoid Linear Unit (SiLU) activation function \citep{silu}. The second group normalization layer is applied to normalize the output from a 3D size 3 convolutional block. Finally, after one more activation function and one 3D size 3 convolutional block, the output from the ``main'' branch is added to the output from the ''skip'' branch. The Up/Down layer denotes the upsampling or downsampling procedure, only applied in \textit{Res Up/Down Blocks}. The decoder part decodes and upsamples the features from the bottleneck and the 'skip' connections from the encoder. The information from the encoder is passed to the decoder at multiple resolution levels, to maintain information at different levels and avoid vanishing gradient issues. In this model, no stochasticity is injected.

The goal of our model is to adjust the fields generated by the stochastic SR model to match as closely as possible the output of a specific HR simulation, conditioned on the HRICs. Due to the limitations of GPU memory, we first split the SR field and HRICs into equal size chunks, then concatenate them in a channel-wise fashion since they have the same spatial dimensions. Both chunks pass through Emulator, and are then compared to the real HR simulations generated from the same HRICs. To evaluate our emulator's output, we use the Charbonnier loss function \citep{Charbonnier}, which is defined as 
\begin{align}
    {L}_{Charbonnier}(x, y) = \sqrt{(x - y)^2 + \epsilon^2}
\end{align}
Here, $\epsilon$ acts as a hyperparameter regulating the tolerance of the loss function. We evaluate our output simulation using a combination of multiple objective functions. The simplest loss function aims to minimize the errors in the displacement and velocity fields in a Lagrangian description: 
\begin{align}
    \mathcal{L}_{\mathrm{Lag}} = {L}_{Charbonnier}(\mathbf{d}_{output}, \mathbf{d}_{HR})
\end{align}
In Papers I, II and III, we have shown that adding Eulerian density field information greatly improves the model performance. Hence, we calculate $n(\mathbf{x})$, which is the particle number in voxel $\bf{x}$, using a CIC scheme. The second term in our loss function is 
\begin{align}
    \mathcal{L}_{\mathrm{Eul}} = {L}_{Charbonnier}(n(\mathbf{x})_{output}, n(\mathbf{x})_{HR})
\end{align}
Using a discriminator to guide the training process (i.e., through adversarial loss) is a common technique in image generation and image-to-image translation tasks, as it enhances the perceptual quality of the output. \xw{We adopt the same definition of adversarial loss, $\mathcal{L}_{Adv}$ as in previous SR works \citep{AI1, AI2, Ai3}}. The total loss for our emulator network is:
\begin{align}
    \mathcal{L}_{total} =  \mathcal{L}_{\mathrm{Lag}} + \mathcal{L}_{\mathrm{Eul}} + \lambda\mathcal{L}_{\mathrm{Adv}}
    \label{eq:loss_function}
\end{align}
The parameter $\lambda$ serves as a weight to balance the contributions of the different loss terms, and its value is empirically set to $3\times 10^{-2}$. This value was determined through experimentation to achieve an optimal trade-off between accuracy and perceptual quality in the outputs.

\subsection{Training Specifics}
\label{subsection2.4:setup}
We set the base channel of our U-net to 64, with channel multipliers at different resolutions set to [1, 1, 2, 2] and attention at the last level. For the discriminator network, the base channel is 32 and the channel multipliers at each resolution are [1, 2, 4, 8]. The Eulerian field, sampled on meshes of side length 1024 cells, is concatenated with a 6D field of particle displacements and velocities, serving as the input for our discriminator. Data augmentation techniques are applied, including random flips along three different axes and random shifts of the chunk crop position. \xw{The input into the Emulator consists of overlapping chunks of size 128, with the overlap region determined by random shifts in the chunk crop position. Periodic boundary condition is applied during the random cropping. }\xw{We use Rectified Adam \citep{RAdam} during training to avoid identified convergence problem suffered by the Adam optimizer, which may converge to bad/suspicious
local optima. We set parameters $\beta_1 = 0.9$ and $\beta_2 = 0.999$ which is the default in Pytorch \citep{pytorch}. The batch size is set to 1 which is the largest size possible given our setup limited by the GPU memory}. Training was conducted using NVIDIA A100 GPUs.

%% file: Sec3_Result1_visual.tex
\section{Results}
\label{section3:Results}
We now compare the outputs of our Emulator model with full N-body HR runs. In our previous work on stochastic SR generations \citep{AI1, AI2, Ai3}, we evaluated results using various summary statistics, including power spectra and halo mass functions. \xw{In this study, because we emulate exact HR fields, we are able to perform particle-level comparisons rather than relying solely on statistical measures. To complement this, we incorporate additional analyses, including visual comparisons of individual structures, evaluation of prediction errors for particle displacement and velocity, and cross-correlation of individual Fourier modes between the emulated and HR runs. Furthermore, we compare the halo-level properties of the emulated and HR simulations by analyzing merger trees and mass histories.}

\subsection{Direct comparison}
\label{subsection3.1: Visual}

\begin{figure*}
\centering
  \includegraphics[width=0.99\textwidth]{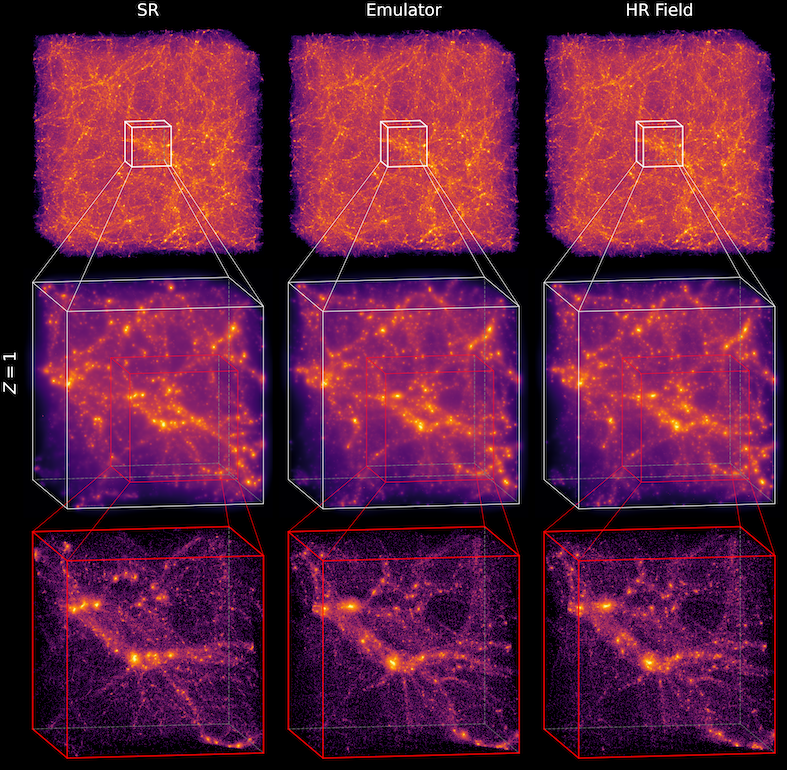}
  \caption{\xw{3D visualization of the SR, Emulator, and HR dark matter density fields with boxsize $100 \mpch$ (first row), $20 \mpch$ (second row), $11 \mpch$ (third row) at $ z = 1 $ from same test set.}}
  \label{fig:3d-visual-z1}
\end{figure*}

\begin{figure*}
\centering
  \includegraphics[width=0.99\textwidth]{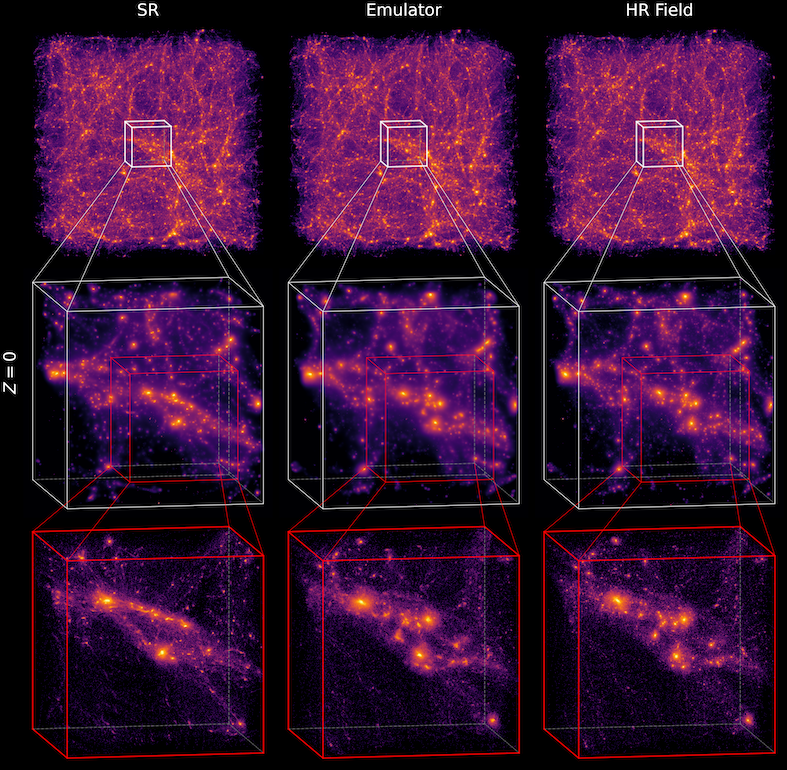}
  \caption{3D visualization of the SR, Emulator, and HR dark matter density fields with boxsize $100 \mpch$ (first row), $20 \mpch$ (second row), $11 \mpch$ (third row) at $ z = 0 $.}
  \label{fig:3d-visual-z0}
\end{figure*}

For our initial validation, we perform a visual analysis of the differences between the HR fields from $N$-body simulations the SR fields generated by our SR model, and the outputs of our newly designed Emulator model. Figures \ref{fig:3d-visual-z1} and \ref{fig:3d-visual-z0} present three-dimensional visualizations of dark matter density plots, created by projecting all particles contained within the box. The images are rendered using the \texttt{gaepsi2} code.\footnote{\url{https://github.com/rainwoodman/gaepsi2}} The first row shows the whole $(100 \mpch)^3$ volume, centered on several massive halos, highlighting the large-scale structures. The second and the third rows focus a sub-volume of $(20 \mpch)^3$ and $(11 \mpch)^3$, respectively, to illustrate finer details and halo distributions.

The visual representation of the density fields reveals that both our SR and Emulator models are capable of generating large-scale structures appearing morphologically similar to one another. On smaller scales, the SR model produces visually authentic features constrained by the LR input, with the fine details appearing reasonable. Our Emulator further refines these small-scale details to better match the HR by conditioning on the HRICs. While the SR model successfully forms halos in regions where the LR input lacks resolution, the Emulator adjusts their positions to align with the HR simulation. For instance, as shown in the bottom row of Figure \ref{fig:3d-visual-z0}, the SR field generates a series of halos lying along a thin filament, the emulator adjusts the field to align with the HR field, placing three massive halos at the center of the box.

\begin{figure}
\centering
  \includegraphics[width=\columnwidth]{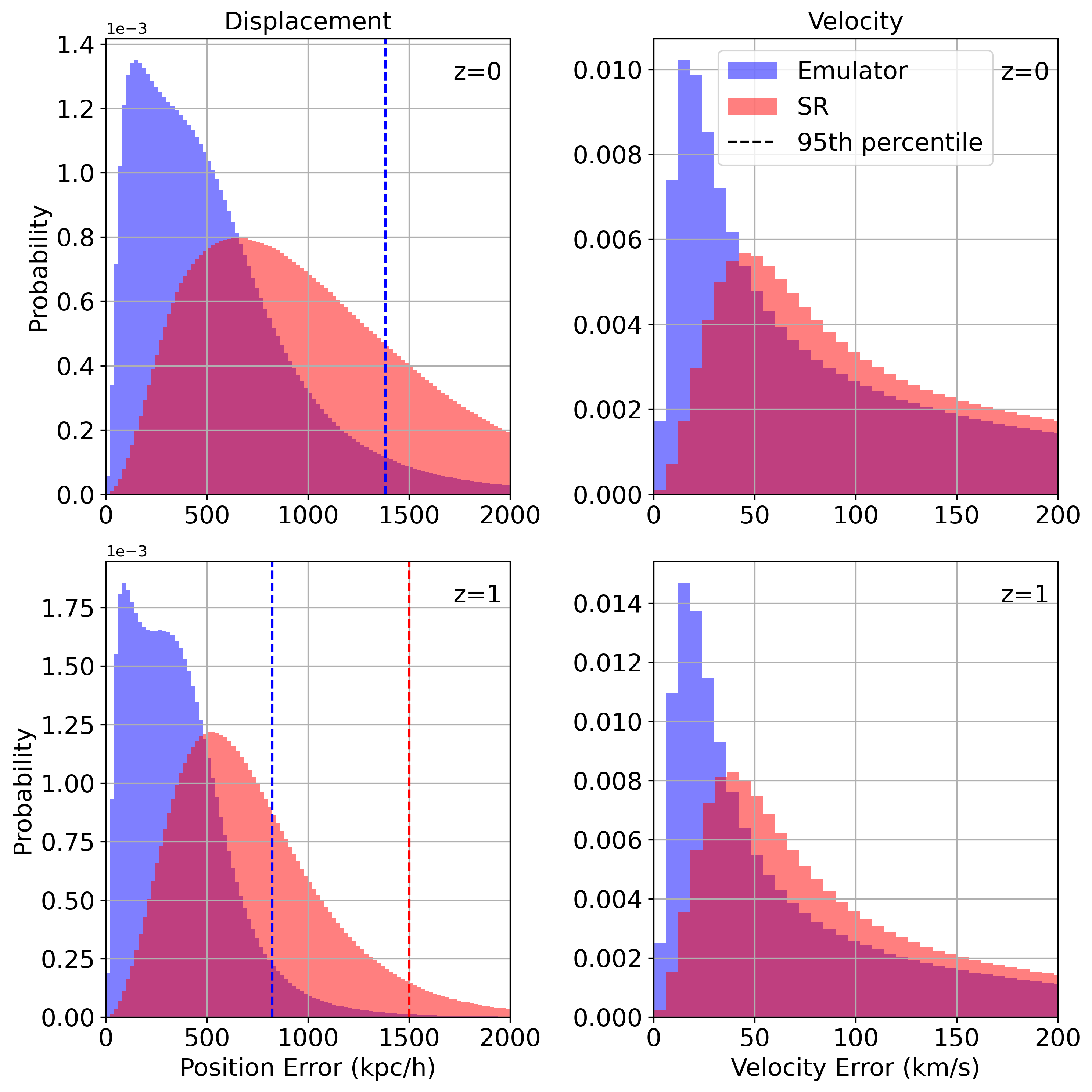}
  \caption{Error distribution of displacement and velocity at $z = 0$ and $z = 1$. The dashed line indicates the 95th percentile error of the Emulator and the SR output.}
  \label{fig:rms}
\end{figure}

\xw{We begin by validating our results using particle-level error. Figure \ref{fig:rms} shows the distribution of root mean squared error distributions for displacement and velocity at $z = 0$ and $z = 1$. The Emulator's error is shown in blue, while the SR error is shown in red. The Emulator model significantly reduces displacement error, with the blue dashed line indicating the 95th percentile error. At $z = 0$, 95 \% of particles from the Emulator have a displacement error below approximately $1.3 \mpch$, while at $z = 1$, this threshold decreases around $0.8 \mpch$. In comparison, the red dashed line indicates the 95th percentile displacement error for the SR output, which is significantly higher than that of the Emulator. Specifically, the SR 95th percentile error is approximately $2.4 \mpch$ at $z = 0$ and $1.5 \mpch$ at $z = 1$, roughly double the Emulator's threshold. Similarly, velocity error distributions also show substantial improvement. The Emulator's 95th percentile velocity error is about $730 \kms$ at $z = 0$ and $460 \kms$ at $z = 1$, whereas the SR velocity error is around $1000 \kms$ $z = 0$ and $630 \kms$ at $z = 1$. These results prove that our Emulator is capable of reducing the stochasticity in the SR model's output, impoving the accuracy significantly. The enhanced alignment of particle positions and velocities with the HR simulation confirms that the Emulator successfully utilize the information from the HRICs to guide the adjustments of the stochastic SR model's output.}

%% file: Sec3_Result2_spectrum.tex
\subsection{Fourier analysis: cross-correlation of individual modes}
\label{subsection3.2: Power Spectra}
The over-density field at position $x$, is defined as 
\begin{align}
    \delta(x) = \frac{\rho(x) - \Bar{\rho}}{\Bar{\rho}}
\end{align}
Where $\rho(x)$ is the density at position $x$ and $\Bar{\rho}$ is the average density. The Fourier transform is used to compute the density modes $\delta(k)$, and hence the usual matter power spectrum is given by:
\begin{align}
    \left\langle\delta(\mathbf{k}) \delta\left(\mathbf{k}^{\prime}\right)\right\rangle=(2 \pi)^3 \delta^3\left(\mathbf{k}+\mathbf{k}^{\prime}\right) P(k)
\end{align}
Here $\delta^3\left(\mathbf{k}+\mathbf{k}^{\prime}\right)$ denotes the 3D Dirac delta function. The Nyquist frequency is $k_{\rm{Nyq}} = \pi N_{\rm mesh} / L_{\rm box}$. In our test set, $N_{\rm mesh} = 512$ and $L_{\rm box} = 100 \mpch$. 

We quantify the difference between the SR prediction and the true HR simulation by calculating the cross-correlation coefficient of individual modes in Fourier space:

\begin{align}
    r(k) &= \frac{\left\langle\delta_{\text {pred}}(\mathbf{k}) \delta_{\text {true}}\left(\mathbf{k}^{\prime}\right)\right\rangle}{\sqrt{\left\langle\delta_{\text { pred }}(\mathbf{k}) \delta_{\text { pred }}\left(\mathbf{k}^{\prime}\right)\right\rangle\left\langle\delta_{\text { true }}(\mathbf{k}) \delta_{\text { true }}\left(\mathbf{k}^{\prime}\right)\right\rangle}} \\
    &= \frac{P_{\text{pred}, \text{true}}(k)}{\sqrt{P_{\text{pred}, \text{pred}}(k) P_{\text{true}, \text{ture}}(k)}}, \label{eq:cross}
\end{align}

where $\delta_{\text{pred}}(\mathbf{k})$ is the predicted density field constructed from the emulator, and $\delta_{\text{true}}(\mathbf{k})$ is the density field constructed from the real HR N-body simulations. Here, $r(k)$ measures the correlation between the phases of the modes of the two fields. In an ideal scenario, where the true and predicted fields are perfectly correlated, the cross-correlation between the two fields would be 1 across all scales. The stochasticity is defined as $1 - r^2(k)$, which quantifies the variation in the prediction that cannot be accounted for by the HR simulations. 

\begin{figure*}
\centering
  \includegraphics[width=\textwidth]{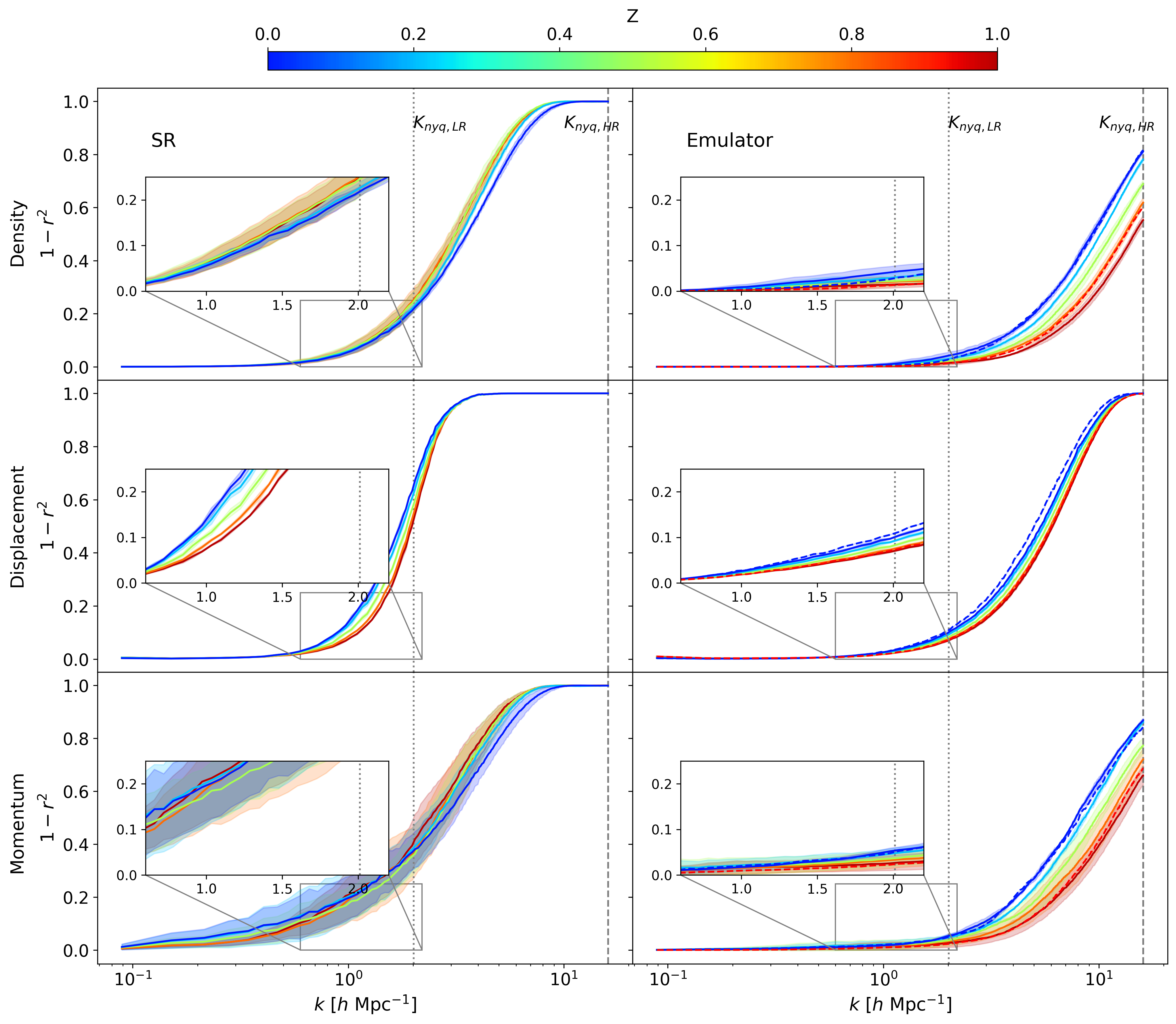}
  \caption{Stochasticity, derived from the correlation coefficient of individual Fourier modes as a function of wavenumber. From left to right, we present results for the cross-correlation of HR (the true N-body simulation) with SR and the Emulator output. The color of each curve corresponds to the redshift, as indicated by the color bar. The shaded area shows the $1\sigma$ standard deviation calculated from all test sets. The vertical dotted gray line is the Nyquist frequency of the LR simulation and the dashed gray line indicates the Nyquist frequency of HR simulation.}
  \label{fig:cross_spec}
\end{figure*}

To quantify agreement of the velocity field, we examine the momentum in Fourier space. The momentum field $p(x)$ is defined as momentum at position $x$. This is calculated using the same \xw{Cloud In Cell (CIC)} scheme as the matter density field. The Eulerian momentum power spectrum is given by:
\begin{align}
    \left\langle p(\mathbf{k}) p\left(\mathbf{k}^{\prime}\right)\right\rangle=(2 \pi)^3 \delta^3\left(\mathbf{k}+\mathbf{k}^{\prime}\right) P_{momentum}(k)
\end{align}
The momentum cross-correlation coefficient is calculated using the analogous equation to Eq. \ref{eq:cross}. Since each particle is initially assigned to a uniform grid with the same number of cells as our density and momentum field, we can also work with the displacement field in Fourier space. The displacement power spectrum (on the initial grid) is given by:
\begin{align}
    \left\langle d(\mathbf{k}) d\left(\mathbf{k}^{\prime}\right)\right\rangle=(2 \pi)^3 \delta^3\left(\mathbf{k}+\mathbf{k}^{\prime}\right) P_{displacement}(k)
\end{align}
We use the same definition as Eq. \ref{eq:cross} for the cross-correlation. The results are plotted in Figure \ref{fig:cross_spec}, with the color of each curve representing the redshift of its corresponding simulation. We select 5 snapshots from simulation at redshift $z = 1, 0.8, 0.5, 0.2, 0$. 

The results of the density power spectrum error are presented in the top row of Figure \ref{fig:cross_spec}. The outputs of the Emulator model are considerably more accurate than those of the SR model. At $k = 1 \invhmpc$, near the LR Nyquist frequency, the SR density field deviates from the HR model with a stochasticity of approximately 7\%, whereas our emulator maintains less than 2\% across all redshifts. The discrepancy between the SR and HR fields increases to over 99\% stochasticity at $k = 8.5 \hmpc$, but the Emulator reduces this to around 50\%. The error at the HR Nyquist frequency is also reduced from 100\% for SR to between 55\% and 80\% for the Emulator, depending on redshift. Notably, a significant fraction of the structures are reproduced even at $k = 10 \invhmpc$ and beyond. 

The second and third rows show the error in the displacement and momentum power spectra, respectively. We observe significant improvements in accuracy with the Emulator model. The displacement power spectrum error in the SR simulation, which is around 60\% near the LR Nyquist frequency, is reduced to less than 10\% by our Emulator. Regarding the momentum power spectrum error, the Emulator reduces the SR simulation's error from 40\% to less than 5\% near the LR Nyquist frequency. 

\xw{We also trained our Emulator at a single redshift. The blue dashed line and red dashed line in the Emulator panels represent the single-redshift Emulator at $z = 0$ and $z = 1$, respectively. The single-redshift version demonstrates comparable performance to the multi-redshift model, indicating that our multi-redshift approach experiences almost no performance loss compared to the single-redshift model. We also trained the Emulator without the adversarial loss at both single and multiple redshifts; see Appendix \ref{appendix} for details. As shown in Figure \ref{fig:appendix-cross-spec}, we would like to mention the dashed blue lines in the first row, represents the results of the Emulator trained on a single redshift ($z = 0$). This single-redshift version shows a significant improvement compared to the multiple-redshift model, reducing the error at the HR Nyquist frequency to around 12\%, and the error at $k=10 \invhmpc$ to only 5\%. These results represent the best performance the Emulator w/o DG can achieve. However, we have not yet been able to reproduce this level of accuracy when training the Emulator w/o DG across multiple redshifts. While our emulator in its current state is aiming for use cases currently carried out by approximate methods, it is interesting to note that a 5\% error in Fourier space at $k=10 \invhmpc$ is comparable to many full N-body simulation codes. (\xw{see e.g., Figure 13 of \citealt{heitmann08}}).}

We can ask what limits the Emulator's accuracy, and how this can be improved: the movements of particles within a virialized halo are rapid and can vary significantly over time. This high frequency behavior diminishes the Emulator's ability to accurately match the positions and velocities of particles at the particle level. This difficulty is further compounded by the fact that HRICs do not contain information from non-linear regimes. Additionally, due to memory constraints, the full simulation must be split into cubical chunks, with the GPU handling a maximum of $128^3$ particles per chunk. A single large halo may span multiple chunks, meaning the emulator lacks access to all the relevant particle information. This limitation could potentially be mitigated by using a GPU with more memory or use a more memory efficient model. Despite these challenges, the Emulator demonstrates strong performance, effectively adjusting the small-scale details generated by the SR model starting from $k = 1 \invhmpc$, achieving excellent accuracy up to the Nyquist frequency of the initial grid. It also maintains decent accuracy up to $k = 10 \invhmpc$.

%% file: Sec3_Result3_halo.tex
\subsection{Halo catalog analysis}
\label{subsection3.3: halo-catalog}
Dark matter, under gravitational clustering, forms high-density, virialized structures and local maxima within them, known as halos and subhalos. Understanding these structures and substructures that potentially host galaxies is crucial for connecting theoretical models to observations. These halos and subhalos, identified using halo finder algorithms, play a key role in constructing mock observational catalogs from N-body simulations. We use the {\small{SUBFIND}} code \citep{subfind} to determine the positions and masses of halos and subhalos, as well as to build merger trees. {\small{SUBFIND}} identifies substructures by detecting locally dense and gravitationally bound groups of particles. \xw{It starts with the Friends-of-Friends algorithm, followed by the estimation of each particle's local density via adaptive kernel estimation.}

We apply the {\small{SUBFIND}} algorithm to test sets from SR output, Emulator output and HR output of the $N$-body simulation, comparing statistical properties of the halo and subhalo populations. Throughout this analysis, we focus on halos and subhalos containing at least 100 tracer particles in our test sets.

\subsubsection{Halo abundance}
\label{subsection3.3.1: hmf}
The most straightforward comparison involves counting the number of halos as a function of mass. The halo mass  The halo mass function is defined as:
\begin{align}
    \Phi = dn/d\log_{10}{M_h},
\end{align}
where $dn$ is the comoving number density of halos within an infinitesimal logarithmic mass bin $d\log_{10}{M_h}$, and the fractional error is calculated using:
\begin{align}
    error = \frac{\Phi_{Pred}}{\Phi_{HR}} - 1.
\end{align}

\begin{figure*}
\centering
  \includegraphics[width=\textwidth]{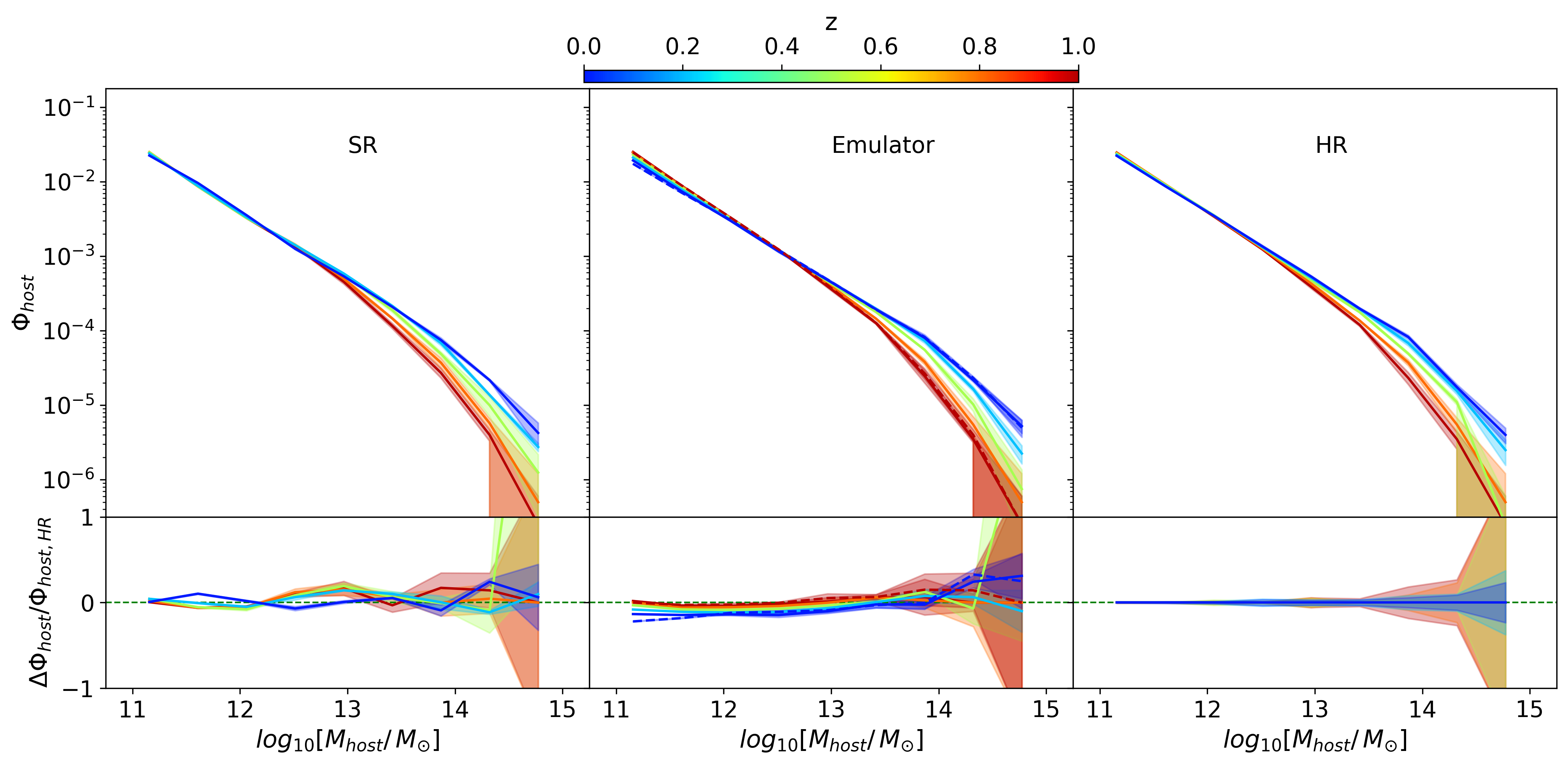}
  \caption{The halo mass function at redshifts $z = 1, 0.8, 0.5, 0.2, 0$ measured from the SR, Emulator and HR fields. The color of each curve denotes the redshift of simulation in test set. The shaded areas indicate $1\sigma$ standard deviation from all test sets. The upper panel shows halo mass functions and lower panel shows the correspondingfractional errors.}
  \label{fig:host_hmf}
\end{figure*}

\begin{figure*}
\centering
  \includegraphics[width=\textwidth]{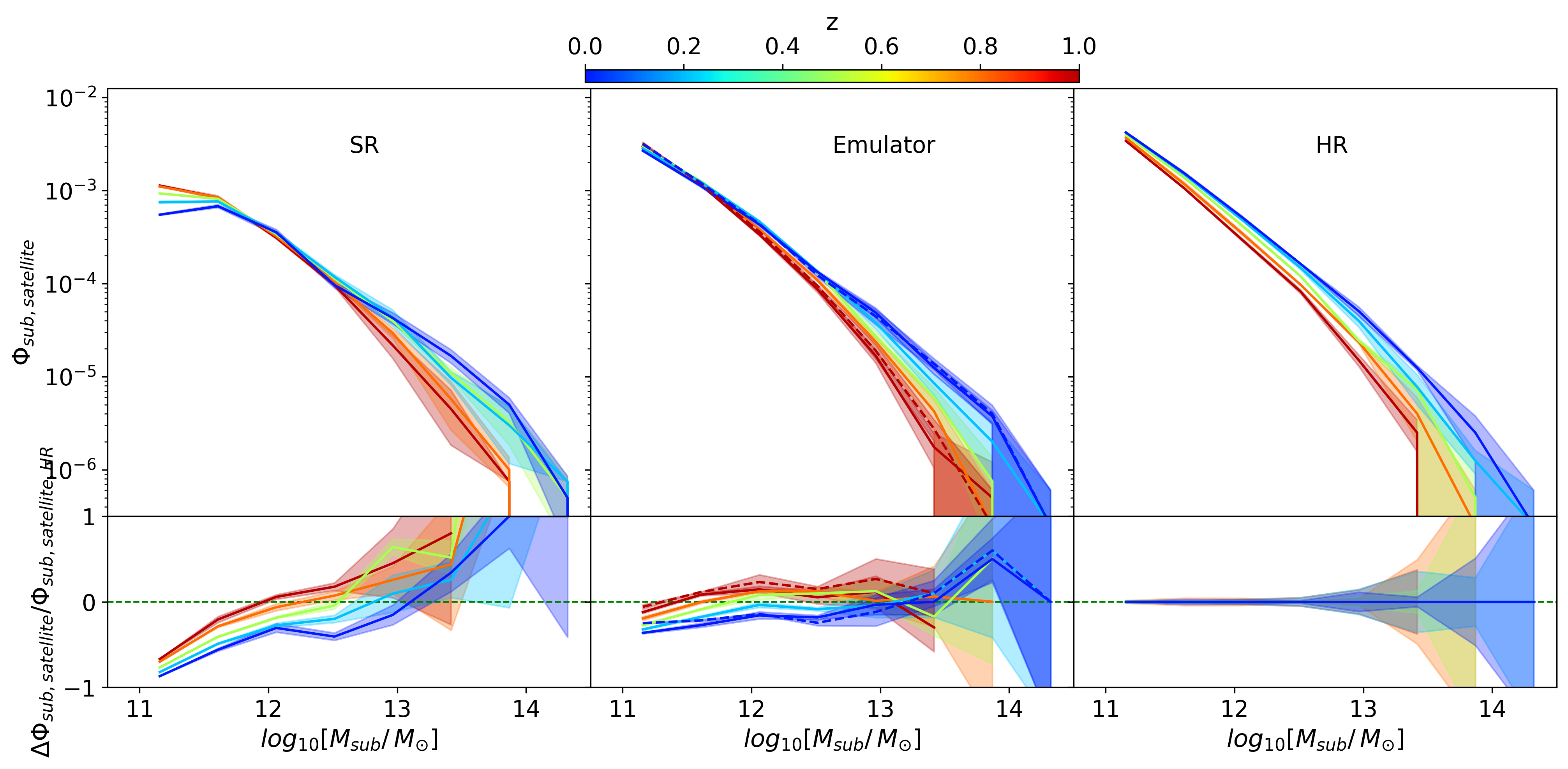}
  \caption{The subhalo mass function at redshifts $z = 1, 0.8, 0.5, 0.2, 0$ calculated from the SR, Emulator and HR fields. The color of each curve denotes the redshift of simulation in test set. The shaded areas indicate $1\sigma$ standard deviation from all test sets. The upper panel shows halo mass functions and lower panel shows the corresponding fractional errors.}
  \label{fig:sub_hmf}
\end{figure*}

Figure \ref{fig:host_hmf} and Figure \ref{fig:sub_hmf} show the mass function for halos and subhalos measured from the SR, Emulator, and HR outputs at five redshifts from $z = 1$ to $z = 0$ from the test sets. The colors of the curves represent the redshift of each simulation, with shaded regions indicating the $1\sigma$ standard deviation. In Figure \ref{fig:host_hmf}, the top panel shows the halo mass function, while the bottom panel presents its fractional error. Similarly, Figure \ref{fig:sub_hmf} illustrates the subhalo mass function in the same format. 

\xw{As shown in Figure \ref{fig:host_hmf}, we observe that the Emulator is biased to lower number densities at lower masses, particularly at $z = 0$. This bias arises because HRICs lack information about small mass halos, hence the Emulator can not accurately capturing the correct abundance of low-mass halos. While the combination of adversarial loss helps mitigate this issue, it does not completely resolve it. As shown in Figure \ref{fig:appendix_halo} ,the absence of guidance from the discriminator exacerbates this bias, resulting in even larger discrepancies.}

Regarding substructures, Figure \ref{fig:sub_hmf} shows that our Emulator model statistically improves over the SR model in the mass range covering $[10^{11}, 10^{13}] \msun$. The SR model exhibits an approximately 90\% deficit in the subhalo mass function at the lower mass end, whereas our Emulator significantly reduces this error by at least 50\%. 

\begin{figure*}
\centering
  \includegraphics[width=\textwidth]{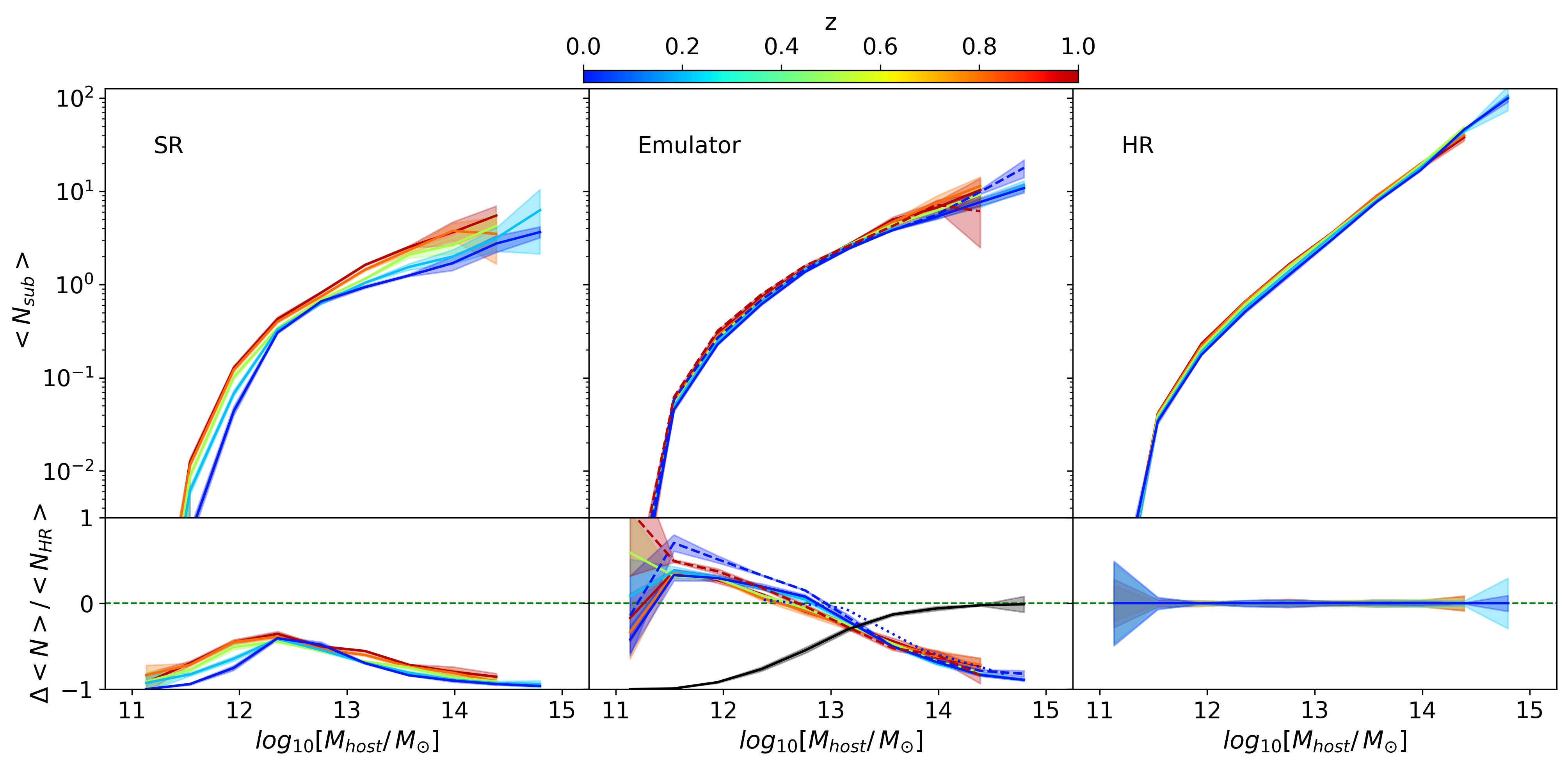}
  \caption{The mean occupation number of subhalos vs host halo mass at redshift $z = 1, 0.8, 0.5, 0.2, 0$ measured from the SR, Emulator and HR fields. The $y$ axis is the averaged number of subhalos in each host halo mass bin. The color of each curve corresponds to its redshift according to the color bar at the top. The shaded area shows the $1\sigma$ standard deviation from all test sets. The dashed blue line represents the mean occupation number at $z = 0$, calculated by selecting only those halos where 99 percent of their particles lie within a single particle chunk (see text). The solid black line in the lower panel of the Emulator represents the average number of subhalos below the Eulerian resolution of the model, shifted down by 1 for clarity (see text for a detailed description).}
  \label{fig:mean_occup}
\end{figure*}

\begin{figure*}
\centering
  \includegraphics[width=\textwidth]{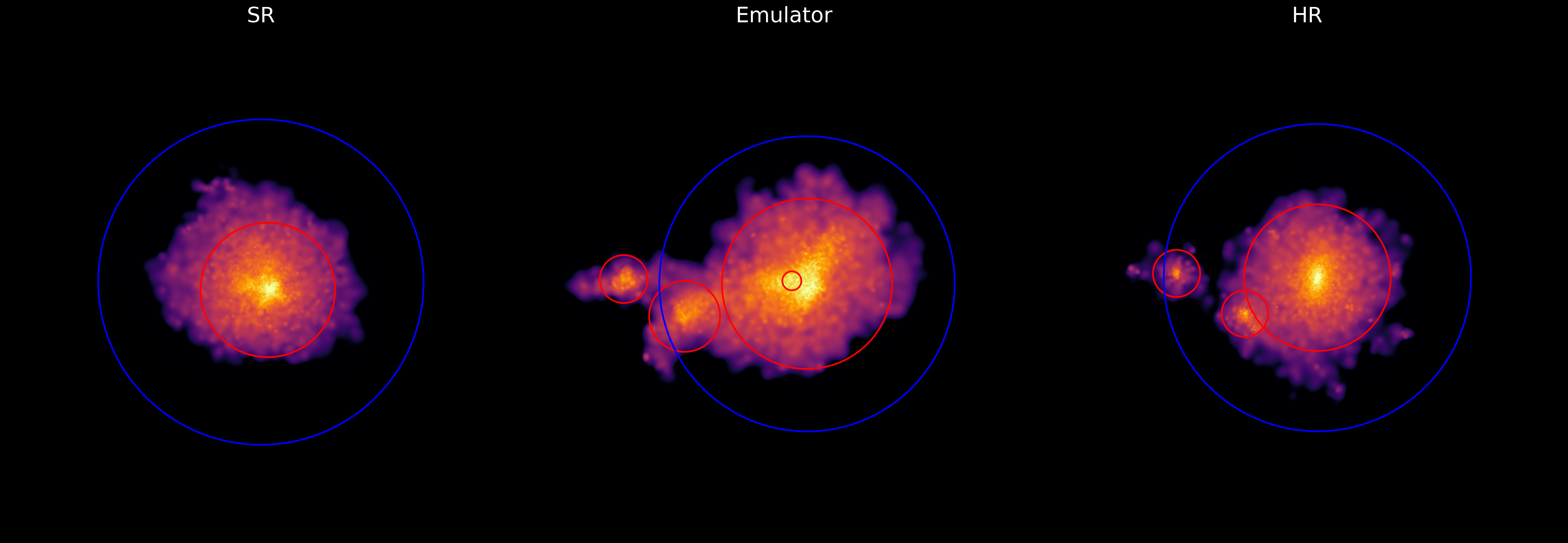}
  \caption{Visual comparison of subhalos from SR, Emulator and HR output, centered on a massive halo with halo mass $1.1*10^{13} \msun$. We mark host halos with blue circles and subhalos with red circles. The radius of the circle is proportional to the radius. All halos shown here have mass $> 100 M_{DM}$.}
  \label{fig:subhalo_1e13}
\end{figure*}

Figure \ref{fig:mean_occup} shows the mean occupation number of subhalos as a function of host halo mass. The SR field exhibits a significant offset from the HR value, predicting $50\% - 90\%$ fewer subhalos per halo compared to the HR simulation. The Emulator improves on this from the lower mass end up to $10^{13} \msun$, reducing the error to between 16\% and 75\%. \xw{As shown in Figure \ref{fig:subhalo_1e13}, for a host halo with a mass of approximately $10^{13} \msun$, the SR output fails to accurately predict the correct number of subhalos. However, the Emulator successfully predicts a subhalo count consistent with the HR simulation and captures their relative positions.} \xw{Both the SR and Emulator's outputs struggle to generate sufficient detail, particularly in highly dense regions. As a result, some smaller subhalos are connected by the halo finding algorithm, forming larger subhalos, which leads to an overabundance of massive subhalos and a deficit at the lower-mass end. The Emulator mitigates this overabundance by further refining the SR output. However, one of the very massive subhalo with mass around $10^{14} \msun$ is misclassified as a satellite halo. This misclassification arises because the Emulator's output remains slightly fuzzier than the target HR simulation, as illustrated in Figure \ref{fig:3d-visual-z0} and Figure \ref{fig:3d-visual-z1}. To address this issue, we manually correct the classification of this halo in our halo catalog.}

Both the SR model and the emulator predict fewer subhalos within the most massive halos as shown in Figure \ref{fig:mean_occup}. We believe that this is partly due to the limited Eulerian field resolution within our discriminator. \xw{With the current setup, the input to our discriminator consists of 1024 mesh cells per side of a $100 \mpch$ box, resulting in a resolution of approximately $0.1 \mpch$,per Eulerian field "pixel." This resolution limits the discriminator's ability to accurately detect 
substructures within massive halos. To quantify the number of subhalos that are not fully resolved, we define the distance between two subhalos as the distance between their centers, subtracting the half-mass radius of both subhalos. We identify subhalos in the HR simulation where the distance to the nearest neighboring subhalo is below the resolution limit (less than the size of one pixel in the Eulerian field). The mean number of such unresolved subhalos per host halo is plotted in the lower Emulator panel of Figure \ref{fig:mean_occup} for direct comparison. This average unresolved halo-occupation number reveals that fewer subhalos are fully resolved by the Eulerian field as the host halo mass increases.} Increasing the resolution by a factor of $f$ would expand the Eulerian field by a factor of $f^3$, demanding more GPU memory. Despite these challenges, our emulator remains quite successful at predicting substructures. Across the halo mass range from $10^{11}-10^{15} \msun$, the mean number of subhalos per halos varies by three orders of magnitude, with the emulator providing relatively accurate results.

We also discover that some halos are divided between simulation chunks that are processed separately. This occurs during both the training and testing phases and causes the model to lack complete halo information. That this occurs is due to GPU memory constraints which requires the simulation to be broken up into relatively small chunks. To quantify the effect of this, we select halos where 99\% of their particles are contained within the same chunk. \xw{The blue dashed line in the lower panel of Figure \ref{fig:mean_occup} the Emulator illustrates the result}, showing 10\% improvement in the error margin at approximately $10^{11} \msun$. This highlights the significance of this chunking issue in the SR+Emulator procedure. The problem could be mitigated by using a more memory-efficient model or by splitting the model sequentially across multiple GPUs for training and testing.

\subsubsection{Halo cross correlations}
\label{subsection3.3.2: halo_cross}

\begin{figure*}
\centering
  \includegraphics[width=\textwidth]{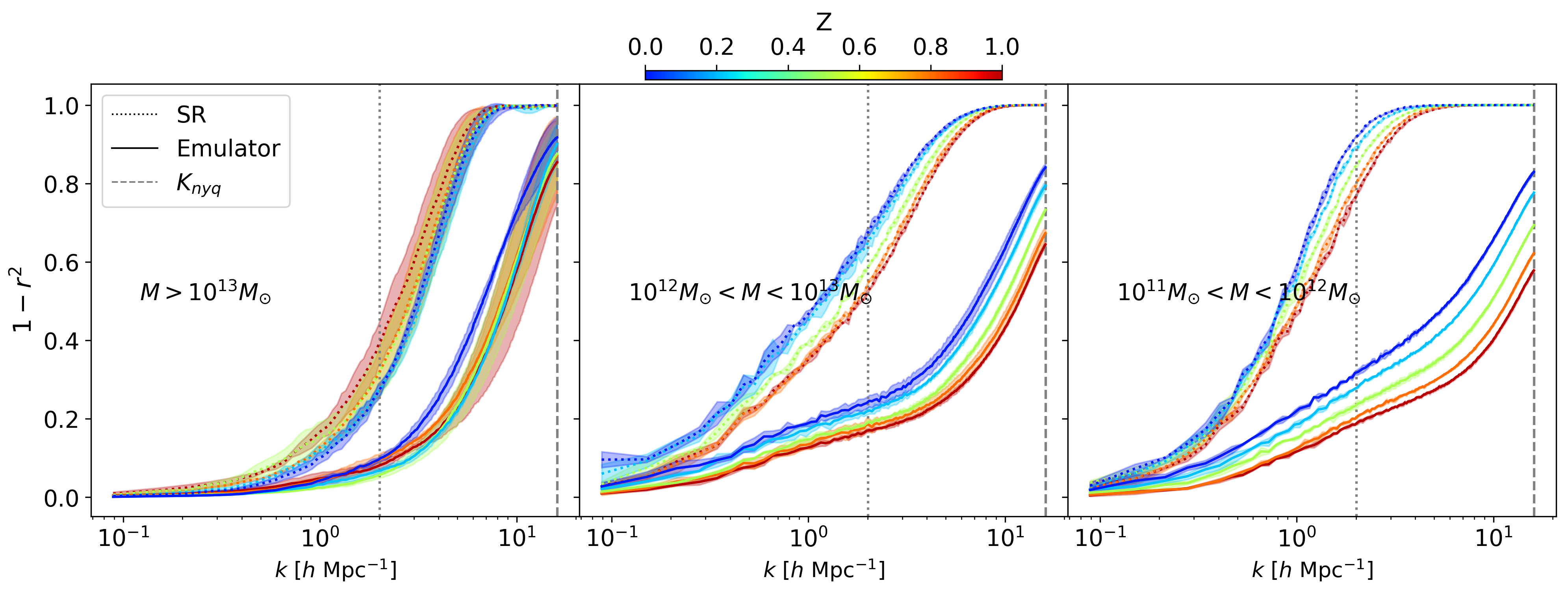}
  \caption{The halo cross correlation error for the SR (dotted lines) and Emulator (solid lines) models for different halo mass regimes (panels from left to right). The color of each curve corresponds to its redshift according to the color bar. The shaded area shows the $1\sigma$ standard deviation from all test sets. The vertical dashed gray lines indicate the HR Nyquist frequency and the dotted gray lines indicate the LR Nyquist frequency.}
  \label{fig:halo_cross_spec}
\end{figure*}

Our Emulator not only refines the SR field but also corrects the positions and velocities of particles, ensuring that the masses and positions of halos are consistent with their counterparts in the HR test sets. To evaluate the accuracy of halo positions and masses, we introduce a new statistic called the "halo cross spectrum". This metric is similar to the cross-correlation coefficient of individual Fourier modes described in Section \ref{subsection3.2: Power Spectra}, but it focuses on halo masses and positions. In this method, each halo is treated as an individual particle, and assigned to a $512^3$ mesh grid using the Cloud-in-Cell (CIC) scheme, with the mass of each particle corresponding to the halo mass. We then perform a Fourier transform on the mesh and deconvolve the CIC window function to obtain the Fourier modes $h(\mathbf{k})$. The power spectrum $P_{h}(k)$ in this case is:
\begin{align}
    \left\langle h(\mathbf{k})) \cdot h(\mathbf{k}^{\prime})\right\rangle=(2 \pi)^3 \delta^3\left(\mathbf{k}+\mathbf{k}^{\prime}\right) P_{h}(k)
\end{align}
The cross-correlation coefficient of individual modes is calculated using the same method applied to the matter density in Section \ref{subsection3.2: Power Spectra}. The results are plotted in Figure \ref{fig:halo_cross_spec}.

Figure \ref{fig:halo_cross_spec} shows the halo stochasticity error. We find that the emulator achieves smaller errors than the SR model across all mass ranges. In the mass range greater than $10^{13} \msun$, the improvements made by Emulator are less significant compared to the enhancements within the mass range of $[10^{11}, 10^{13}] \msun$. This is because our SR model, being conditioned on LR simulations, already constrains the positions and masses of massive halos. The emulator fine-tunes these positions and masses to better match those in the HR field test set. Since the SR model conditions its predictions on LR simulations, it is expected that the correlation between the SR field and the HR field diminishes from $10^{13} \msun$ down to $10^{11} \msun$. However, by adjusting halos based on the HRICs, our emulator achieves significantly higher accuracy within this mass range. In the mass regime below $10^{11} \msun$, the HRICs lack sufficient information on less massive halos, resulting in both the SR field and emulator output failing to predict halos that match the positions and masses of those in the HR field.

\subsubsection{Merger histories}
\label{subsection3.3.3: merger history}
\xw{Given that the Emualtor is capable of reproducing the positions and masses of individual halos}, it is interesting to examine whether the merger histories of these halos are also \xw{consistent with} the HR run. To do this, we use the merger tree algorithm built into the cosmological $N$-body code {\small Gadget-4} \citep{gadget4}, called Hierarchical Bound Tracing (HBT). It uses information from earlier snapshots to help derive the latest halo catalog. Starting from high redshift, the algorithm identifies main halos at their formation and follows the particles within these halos through subsequent snapshots. This process generates a merger tree for the main halos at the first stage, which is then extended to include subhalos. 

\begin{figure*}
\centering
  \includegraphics[width=0.33\textwidth]{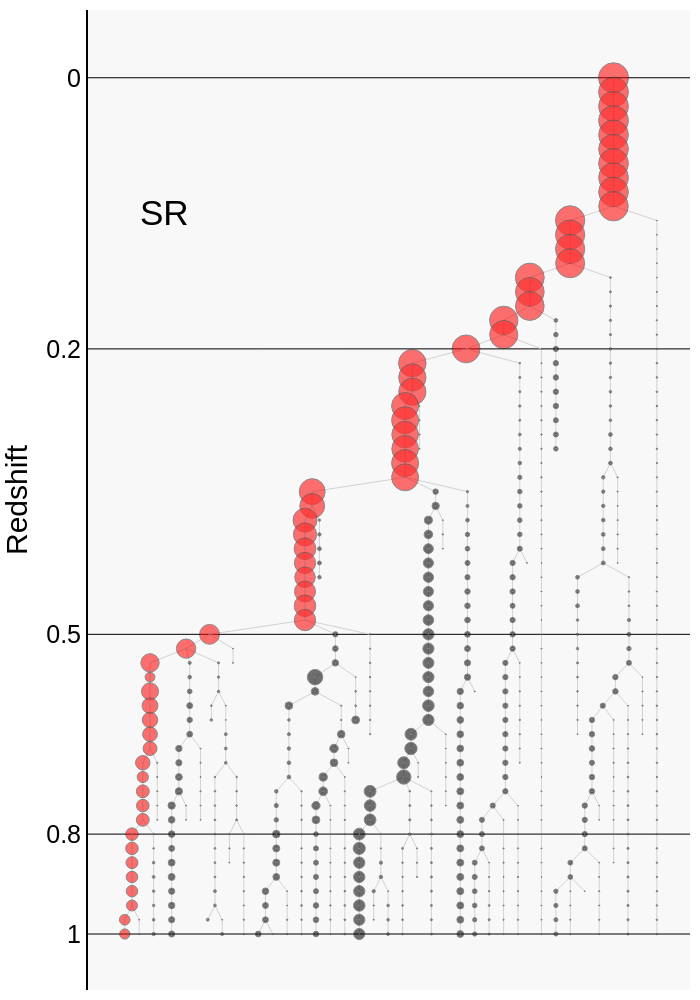}
  \includegraphics[width=0.33\textwidth]{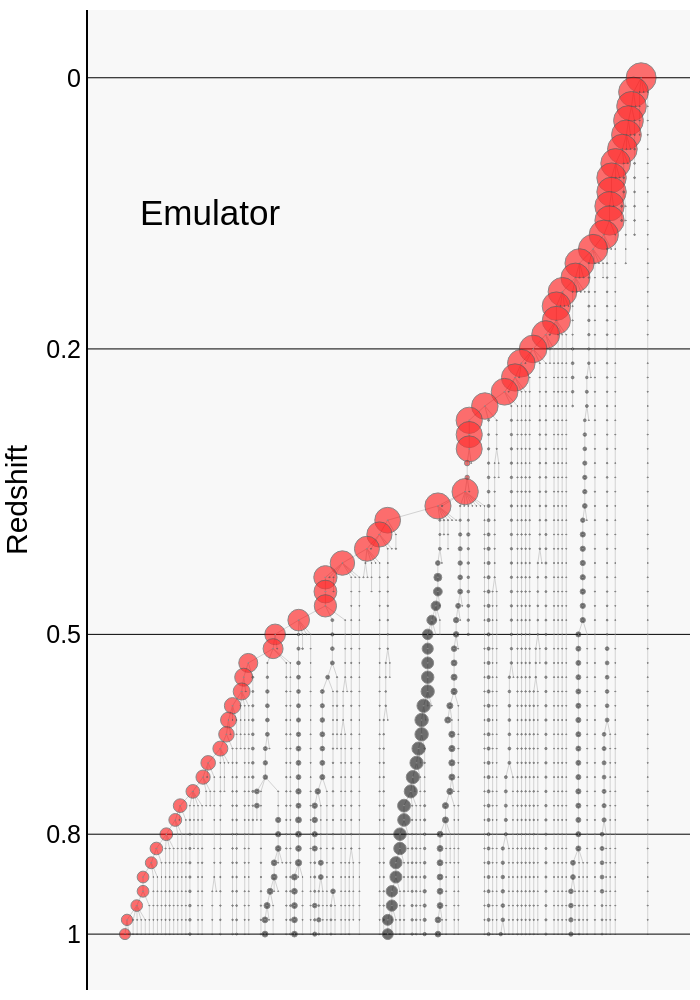}
  \includegraphics[width=0.33\textwidth]{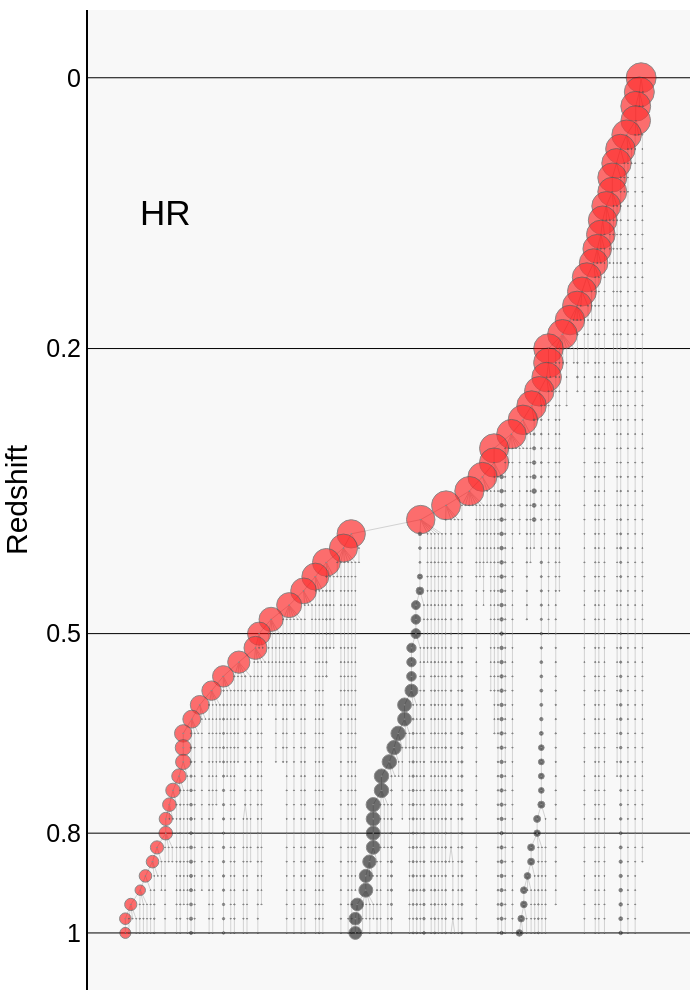}
  \caption{Visualization of merger trees for an example halo in the SR, Emulator and HR field at $z = 0$. Time is depicted as progressing from bottom to top, covering the redshift range of $z = 1$ to $z = 0$. The size of each node is scaled with halo mass. Red circles indicate the main progenitor branch.} 
  \label{fig:merger-tree}
\end{figure*}

Figure \ref{fig:merger-tree} shows a direct visual comparison of merger trees for the same halo selected from the SR field, Emulator output and the HR field (from bottom to top, redshift changes from $z = 1$ to $z = 0$). This halo, with mass $\sim 10^{14} \msun$ was selected based \xw{on the mass relative} to the HR run. The size of each tree node is scaled according to halo mass, with red colors indicating the main progenitors. Although the SR merger tree looks authentic, with mass fluctuations and mergers, the redshifts when the mergers happen and their numbers are very different in detail from the counterpart HR halo. \xw{In contrast, the Emulator's merger tree provides more accurate predictions of the redshift and masses involved in the main merger event. Additionally, the Emulator output includes a greater number of small-mass halos, i.e., more substructures compared to the SR field, and positions them in relatively accurate locations as a function of time.}.

To further validate the actual merger history, such as mass accretion history measured from the SR, Emulator and HR fields, we select halos within different mass ranges from the SR and emulator fields and calculate the mass error relative to the corresponding HR halos from $z = 1$ to $z = 0$. First, we split all the halos in the HR run at $z = 0$ \xw{into three different mass regimes}, then track the mass history of each halo along its main progenitor branch from $z = 1$ to $z = 0$. To demonstrate the accuracy of our emulation and SR model, we calculate the mass error using the following equation:
\begin{align}
    {\rm error} = \frac{M_{\rm halo, pred}}{M_{\rm halo, HR}} - 1
\end{align}

\begin{figure*}
\centering
  \includegraphics[width=\textwidth]{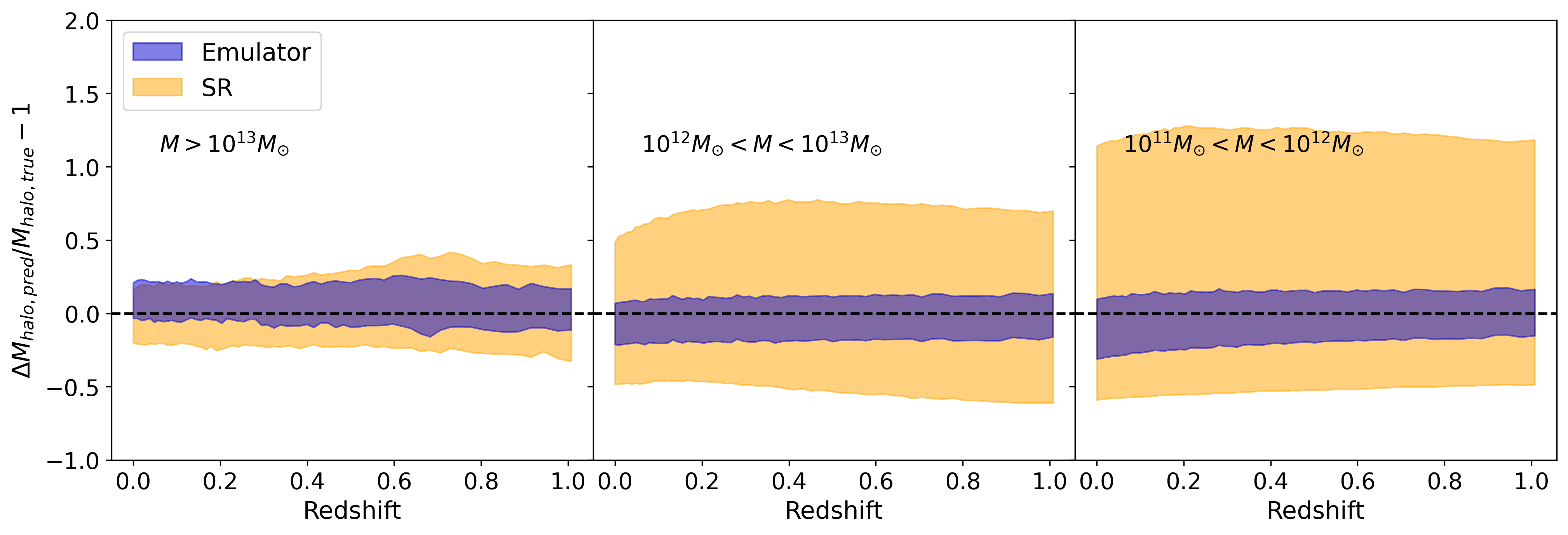}
  \caption{The middle 68\% percentile error regions of halo mass history along the main progenitor branches of halos in the Emulator (blue) and SR (orange) outputs.}
  \label{fig:mah_error}
\end{figure*}

Figure \ref{fig:mah_error} shows the fractional error in the mass history along the main progenitor branch, calculated from one of our test sets. For halos with masses greater than $10^{13} \msun$, our SR model captures the mass to within 40\% error for the middle 68\% percentile, while our Emulator further reduces this margin to within 25\%. As we move to lower mass ranges, $[10^{12}, 10^{13}] \msun$ and $[10^{11}, 10^{12}] \msun$, the Emulator significantly improves the accuracy of the predicted halo mass. In contrast, the SR model, reflecting the stochastic generation of small-scale structures, has a middle 68\% error range $3 \sim 4$ times larger than that of the Emulator.

%% file: Sec4_Discussion.tex
\section{Discussion}
\label{section4:Discussion}

The application of SR techniques has been explored in many different works. Some SR models are designed to generate a range of possible HR simulations based on LR inputs, which could be advantageous when many simulations or mock observations are required. However, this approach can be problematic when a specific HR realization with given initial conditions is needed. The model in our current paper addresses this by incorporating information about the particle initial conditions (displacements and velocities) from the high-resolution $N$-body simulations we are trying to match. In $N$-body simulation algorithms, once the initial conditions are fixed, the simulation results at different redshifts are also deterministic. These initial conditions provide sufficient information to correct the stochastically generated SR field. Combining the SR model with our Emulator, we can generate SR fields with different random seeds and adjust them to the desired results by leveraging the initial displacement and velocity.

The accuracy of the refined SR field, adjusted by our Emulator, represents a significant improvement over the initial SR outputs. This improvement was demonstrated through the analysis of Fourier modes in Section \ref{subsection3.2: Power Spectra}, the halo abundance in Section \ref{subsection3.3.1: hmf} and merger histories in Section \ref{subsection3.3.3: merger history}. As expected, our emulator achieves a much closer correlation with HR field compared to the SR field on scales below the LR resolution, in this case from $k \sim 1 \hmpc$ to the HR model's Nyquist frequency. Additionally, in terms of halo mass function and the mean subhalo occupation number, our Emulator has shown improvement in predicting substructures. We also introduced the "halo cross spectrum", which measures the spatial correlations of halo mass and position, in Section \ref{subsection3.3.2: halo_cross}. This statistic shows that we have improved this correlation. Furthermore, our emulator successfully predicts accurate halo histories, as demonstrated by comparing merger trees and mass accretion histories with the HR runs.

However, there are some discrepancies in agreement with the HR runs. Both the SR model and the Emulator currently exhibit a lack of substructures within very massive halos, particularly at the higher mass end, as seen in Figure \ref{fig:mean_occup}  and the visual comparisons in Section \ref{subsection3.1: Visual}. Most of the missing halos are of low mass, suggesting that increasing the resolution of the Eulerian fields, used to provide direct information about substructures to the discriminator, could improve this. At current spatial resolution, the discriminator struggles to identify substructures within massive halos that require higher contrast to resolve.

Another area for future improvement is related to the need to split the entire $N$-body simulation (with $512^3$ particles) into $128^3$ chunks due to limited GPU memory. This chunking limits the emulator's knowledge to correlations and interactions among particles within the same chunk, making it impossible to identify substructures that span multiple chunks. Increasing the chunk size or the resolution of the Eulerian field is challenging, as an increase by a factor of $f$ leads to $f^3$ times more memory consumption. This limitation motivates us to search for more memory-efficient methods to incorporate information from the entire $N$-body simulation as well as the density field. Both of these challenges could potentially be mitigated by exploring more advanced neural networks or memory-efficient strategies.

\subsection{Running the Emulator output forward with the $N$-body code}

The output of our SR and Emulator runs has the same format as the true N-body simulations. In principle, we can treat them in the same fashion, as we have done for halo finding and calculating merger trees. A particularly interesting test involves passing the emulator output directly to the $N$-body code and running it forward in time. If the emulator produces dynamically correct structures, we expect the model to evolve reasonably well, approximating the outcome of a full HR simulation at later redshifts. We have conducted an initial test of this approach.

\begin{figure}
\centering
  \includegraphics[width=0.99\columnwidth]{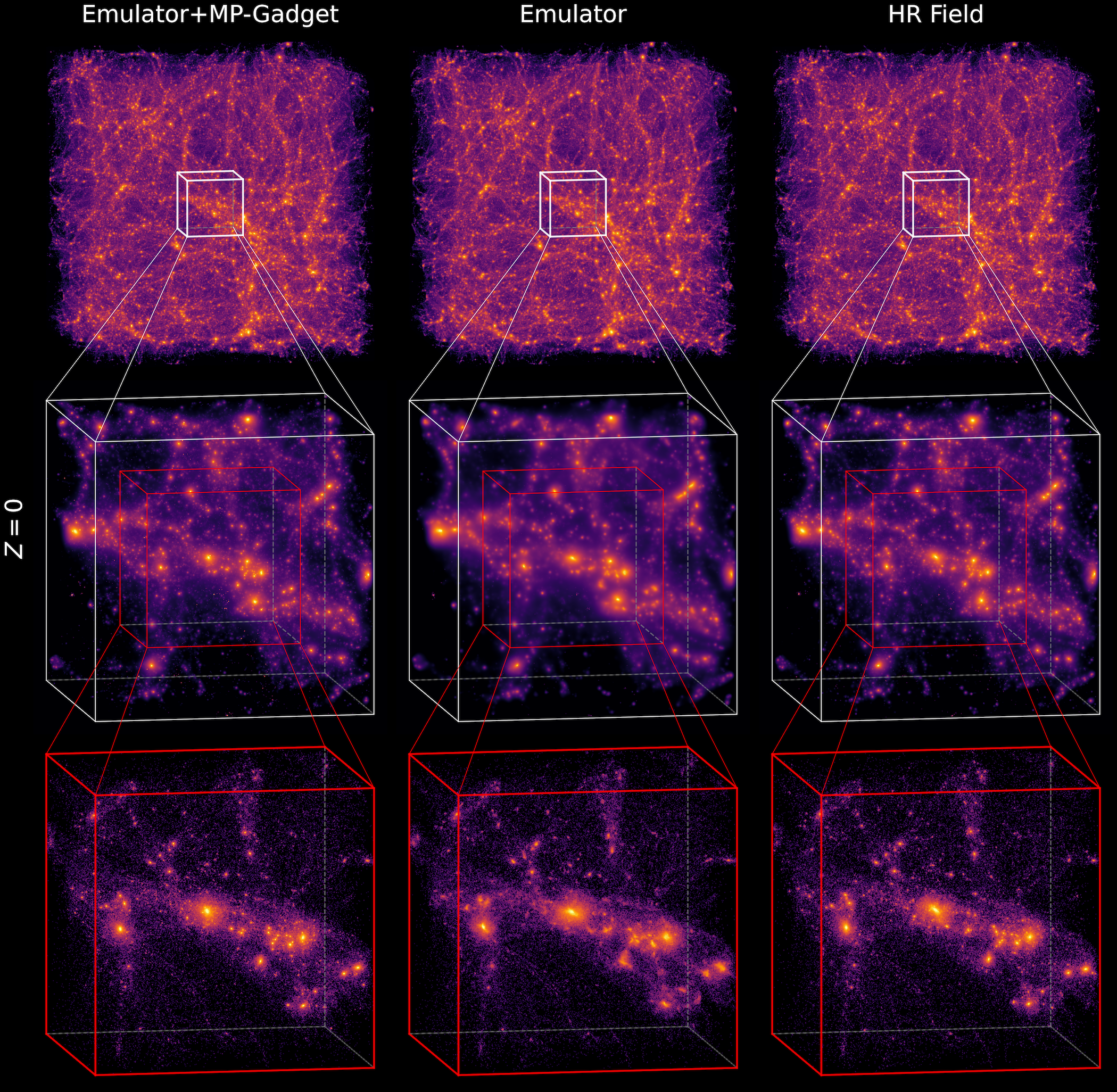}
  \caption{3D visualization of Emulator+MP-Gadget, Emulator model output and HR dark matter density fields with boxsize $100 \mpch$ (first row), $20 \mpch$ (second row), $11 \mpch$ (third row) at $ z = 0 $}
  \label{fig:visual_forward}
\end{figure}

We take the output from our Emulator at an early redshift ($z = 1$) and input it into MP-Gadget to run a forward N-body simulation. We then compare the results of this forward simulation with those of an actual HR simulation. The results, shown in Figure \ref{fig:visual_forward}, demonstrate that this hybrid approach generates more realistic substructures, close to those seen in HR simulations, while still preserving the large-scale structures. This not only validates the accuracy of our model's predictions but also suggests potential for a hybrid N-body+Emulator method. We plan to explore this further in future work.

%% file: Sec5_Conclusion.tex
\section{Summary and Conclusions}
\label{section5:Conclusion}
We have developed and evaluated an emulator based on neural networks that produces super-resolved cosmological $N$-body simulation outputs. The inputs to the Emulator include a low-resolution $N$-body simulation and the initial conditions from a high-resolution simulation. The Emulator refines the outputs of the SR model presented in  \citealt{Ai3}, which is based on the LR simulation. By integrating the initial conditions of particles (including initial displacement and velocity), the Emulator predicts the adjustments to the SR model's 6D phase space output. These adjustments ensure that the emulator's output closely aligns with the HR $N$-body simulation using the same initial conditions. 

We conduct detailed validation of our Emulator by comparing various statistics calculated from test sets. In Section \ref{subsection3.1: Visual}, we present a visual comparison between the SR model, the Emulator and the corresponding HR simulation. These comparisons demonstrate that on large scales, the SR, Emulator and HR fields appear identical. While SR field presents visually convincing small-scale structures, it differs in detail from the HR simulations. However, the Emulator output successfully aligns with the HR simulation even on scales well below the resolution of the LR simulation. We also compared the error distributions between the SR model and the Emulator. The Emulator significantly reduces particle displacement and velocity errors compared to the SR model.

In Section \ref{subsection3.2: Power Spectra}, we analyze the outputs in Fourier space and found that the emulator output correlates much more closely with the HR field than SR output, from $k = 1 \hmpc$ down to the Nyquist frequency. The Emulator's accuracy represents a significant improvement over the SR field. To evaluate the halo population and substructures, we compute the halo and subhalo mass functions, as well as mean occupation number, in Section \ref{subsection3.3: halo-catalog}. Both the SR and Emulator host halo mass abundance have good statistical agreement with HR, showing our Emulator does not diminish the previous results from SR field (\citealt{Ai3}).  At the same time, the subhalo mass function shows that the Emulator output improves significantly over the SR field,  adjusting the SR field to the HR field and improving the prediction of small-scale structures. However, a deficit in the subhalo abundance for halos of mass $[10^{13}, 10^{14}] \msun$ indicates that current model still struggles to predict a sufficient number of substructures within large mass halos, likely due to the limitations of the Eulerian field resolution.

To quantify the correlation of halo properties (position and mass) between the Emulator and HR field, we introduced a "halo cross spectrum" to evaluate this in Fourier space. In Section \ref{subsection3.3.2: halo_cross}, we show that our Emulator output leads to significant improvement over SR in the halo mass regime $ > 10^{11} \msun$. Additionally, in Section \ref{fig:merger-tree}, we validate the Emulator's performance over time by visually comparing the merger trees of a massive halo from the SR, Emulator and HR fields. We find that the Emulator successfully corrects the merger history compared to the SR result.

Our results demonstrate the Emulator's ability to adjust the SR field output to closely match the corresponding HR field by leveraging the initial conditions. The Emulator represents a promising foundation for emulating $N$-body simulations under different constraints, suggesting the feasibility of generally emulating outputs for simulations with varying cosmological parameters across a wide range of redshifts. The integration of the SR model with the Emulator framework enables the rapid and accurate generation of results across various redshifts, providing an effective tool which can compete with other approximate methods such as those of \cite{fastpm} and \cite{pthalos}, while offering greater accuracy and the potential inclusion of more physics. The Emulator approach facilitates the generation of mock halo catalogs and the computation of non-linear statistics, such as power spectrum covariance matrices, positioning it as a valuable tool for constraining cosmological parameters from upcoming cosmological surveys.

%% file: Sec6_Appendix.tex
\section{Results from Training the Emulator Without Discriminator Guidance}
\label{appendix}
We present the results of training the same Emulator without discriminator guidance. In this context, "without discriminator guidance" means training the neural network using the loss function in Equation \ref{eq:loss_function}, but without the third adversarial term.

\begin{figure*}
\centering
  \includegraphics[width=\textwidth]{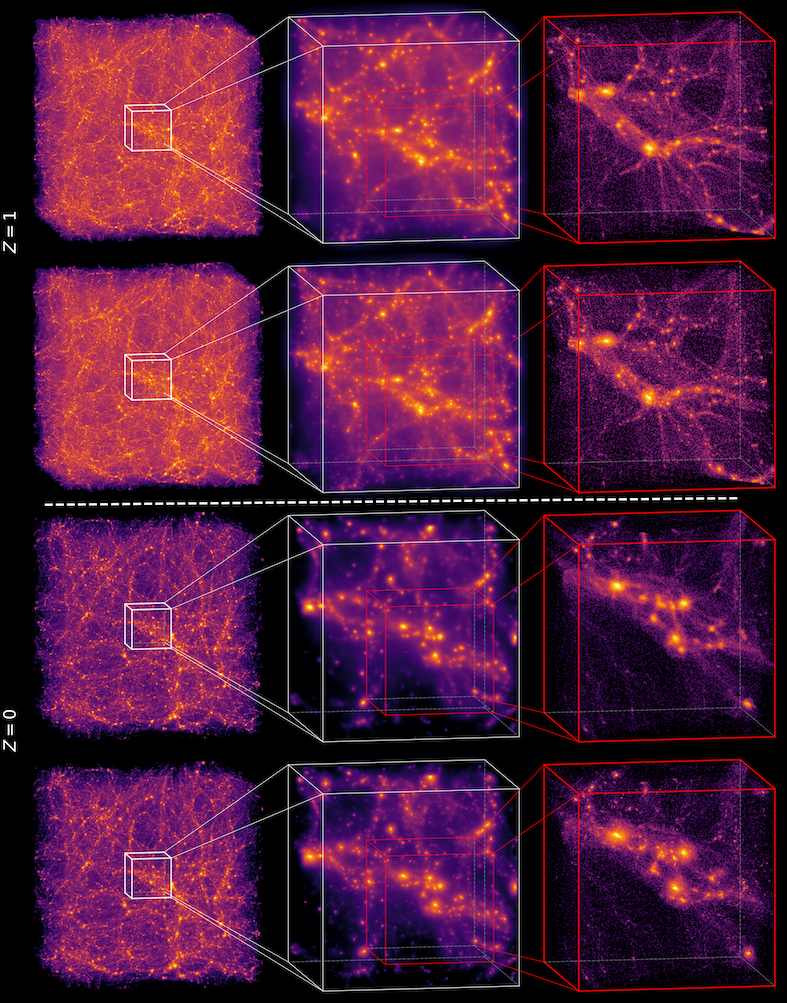}
  \caption{3D visualization of the Emulator w/o DG density fields with boxsize $100 \mpch$ (first column), $20 \mpch$ (second column), $11 \mpch$ (third column) at $ z = 1 $ and $ z = 0 $. The first and third rows correspond to the Emulator w/o DG (without discriminator guidance), while the second and fourth rows represent the standard Emulator.}
  \label{fig:appendix-visual}
\end{figure*}

Figure \ref{fig:appendix-visual} shows the result from Emulator w/o DG model. The first column shows the whole $(100 \mpch)^3$ volume, centered on the same halo as in Figure \ref{fig:3d-visual-z1} and Figure \ref{fig:3d-visual-z0}, to illustrate the large-scale structures, the second and the third columns show a sub-volume of $(20 \mpch)^3$ and $(11 \mpch)^3$, to show the different detailed structures and halo distributions. Comparing with Figure \ref{fig:3d-visual-z1} and Figure \ref{fig:3d-visual-z0}, we can see that the Emulator w/o DG still appears authentic on large scales, similar to our Emulator+DG model. However, as shown in the third column, on smaller scales, the Emulator w/o DG model produces a significantly smoother density field, leading to a noticeable lack of substructures.

\begin{figure*}
\centering
  \includegraphics[width=\textwidth]{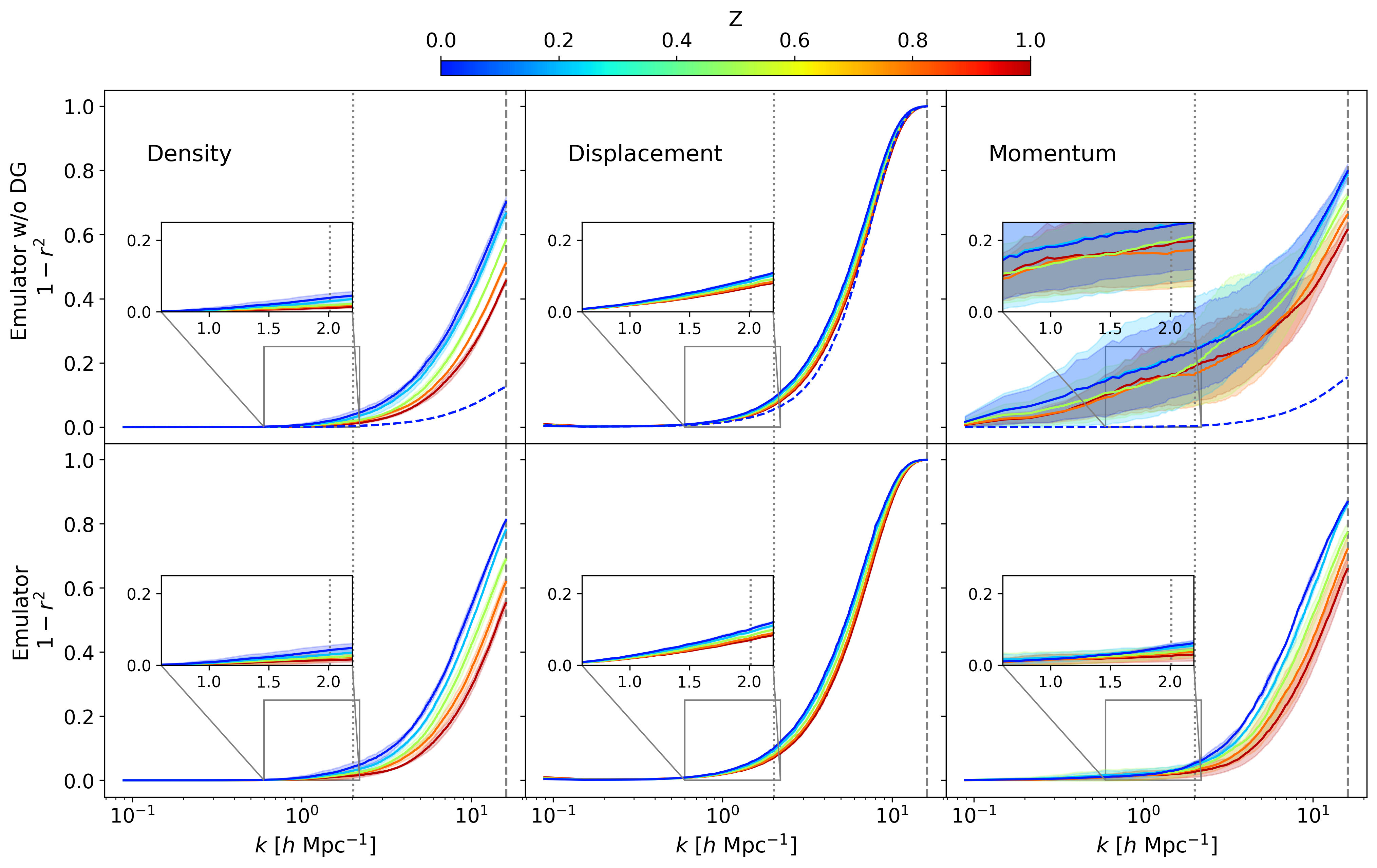}
  \caption{Stochasticity of individual Fourier modes as a function of wavenumber. The color of each curve corresponds to its redshift according to the color bar. The shaded area shows the $1\sigma$ standard deviation from all test sets. The vertical dashed gray line indicates the Nyquist frequency of HR simulation and the dotted gray line is the Nyquist frequency of the LR simulation.}
  \label{fig:appendix-cross-spec}
\end{figure*}

Figure \ref{fig:appendix-cross-spec} shows the results of stochasticity, calculated using the same equations as in Section \ref{subsection3.2: Power Spectra}. We can see that the Emulator w/o DG performs slightly better in the density field and produces comparable results in displacement and momentum. 

\begin{figure*}
\centering
  \includegraphics[width=\textwidth]{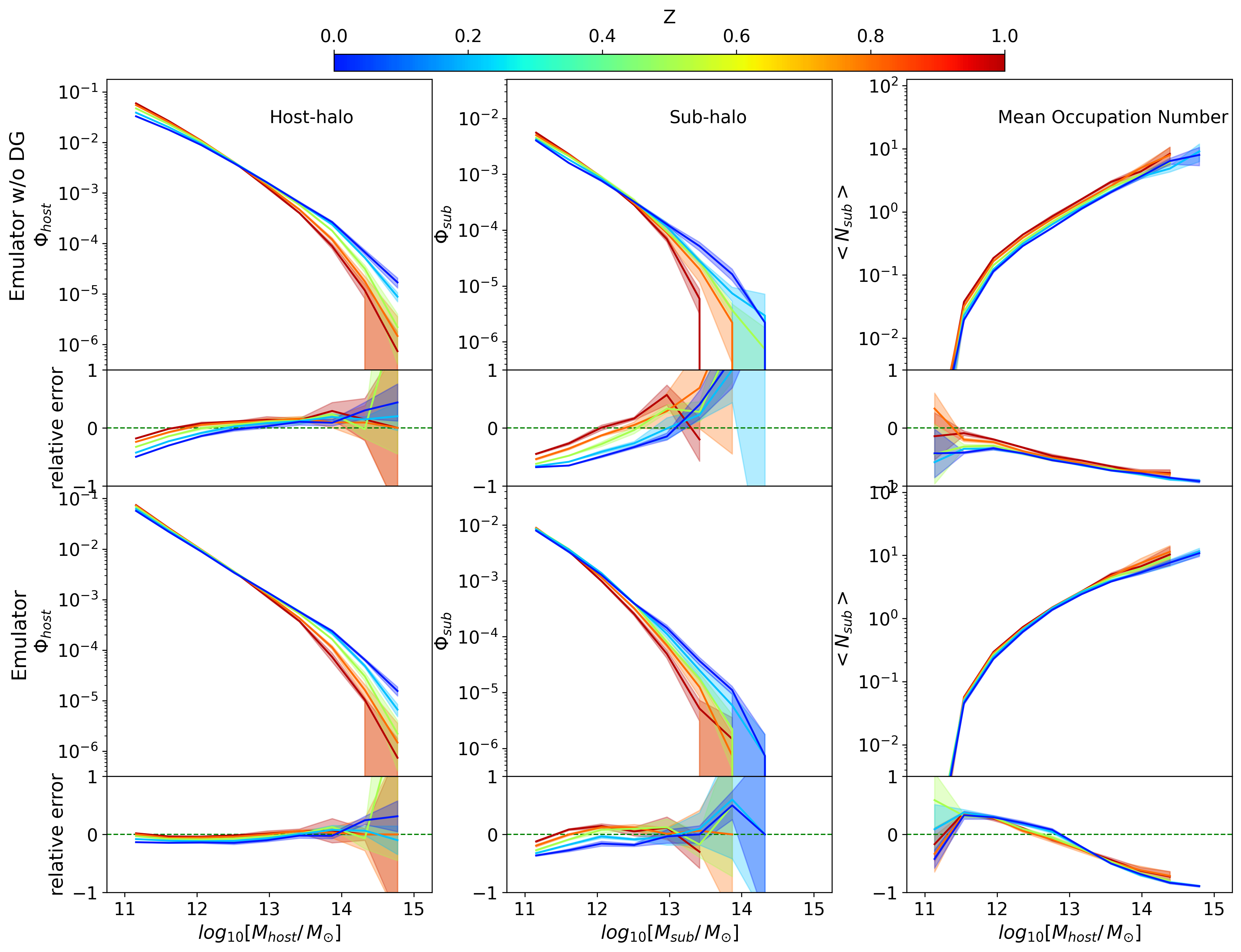}
  \caption{The halo, subhalo mass function and mean occupation number at redshifts $z = 1, 0.8, 0.5, 0.2, 0$ measured from the Emulator w/o DG output. The color of each curve denotes the redshift of simulation in test set. The shaded areas indicate $1\sigma$ standard deviation from all test sets. The upper panel shows halo mass functions and lower panel shows fractional errors}
  \label{fig:appendix_halo}
\end{figure*}

The halo mass function, subhalo mass function, and mean occupation number for the Emulator w/o DG's output are shown in Figure \ref{fig:appendix_halo}. Both the halo and sub-halo mass functions show a deficit in the lower mass region, indicating that training without discriminator guidance (DG) lacks the detailed structures observed on smaller scales in the Emulator+DG model. Additionally, the mean occupation number reveals that training without DG results in fewer low-mass subhalos within the range of $[10^{11}, 10^{13}] \msun$.

Overall we have therefore seen that the discriminator part of the emulator is necessary to produce distinct substructures and a realistic density field. Without it the Emulator does have slightly more accurate Fourier modes but it has a significantly smoother density field on halo scales. We therefore have decided to include the discriminator in the fiducial version of the emulator (used in the main body of the paper).

